\documentclass[sigconf]{acmart}
\usepackage{textcomp}
\usepackage{stfloats}
\usepackage{url}
\usepackage{verbatim}
\usepackage{amsthm}
\usepackage{subfigure}
\usepackage{makecell}
\usepackage{multirow}
\usepackage{enumitem}
\usepackage{subcaption}

\usepackage{algorithm}
\usepackage{algorithmic}

\usepackage[inkscapelatex=false]{svg}
\AtBeginDocument{%
  }

\copyrightyear{2024}
\acmYear{2024}
\setcopyright{acmlicensed}\acmConference[CIKM '24]{Proceedings of the 33rd ACM International Conference on Information and Knowledge Management}{October 21--25, 2024}{Boise, ID, USA}
\acmBooktitle{Proceedings of the 33rd ACM International Conference on Information and Knowledge Management (CIKM '24), October 21--25, 2024, Boise, ID, USA}
\acmDOI{10.1145/3627673.3679697}
\acmISBN{979-8-4007-0436-9/24/10}

\begin{document}

\title[AlignGroup: Learning and Aligning Group Consensus with Member Preferences for Group Recommendation]{AlignGroup: Learning and Aligning Group Consensus with Member Preferences for Group Recommendation}


\author{Jinfeng Xu}
\email{jinfeng@connect.hku.hk}
\affiliation{%
  \institution{The University of Hong Kong}
  \city{HongKong SAR}
  \country{China}}

\author{Zheyu Chen}
\email{zheyu.chen@connect.polyu.hk}
\affiliation{%
  \institution{The Hong Kong Polytechnic University}
  \city{HongKong SAR}
  \country{China}}

\author{Jinze Li}
\email{lijinze-hku@connect.hku.hk}
\affiliation{%
  \institution{The University of Hong Kong}
  \city{HongKong SAR}
  \country{China}}

\author{Shuo Yang}
\email{shuoyang.ee@gmail.com}
\affiliation{%
  \institution{The University of Hong Kong}
  \city{HongKong SAR}
  \country{China}}

\author{Hewei Wang}
\email{heweiw@andrew.cmu.edu}
\affiliation{%
    \institution{Carnegie Mellon University}  
    \city{Pittsburgh, PA}   
    \country{USA}}

\author{Edith C. H. Ngai}
\email{chngai@eee.hku.hk}
\affiliation{%
  \institution{The University of Hong Kong}
  \city{HongKong SAR}
  \country{China}}
\authornote{Corresponding author.}

\renewcommand{\shortauthors}{Jinfeng Xu et al.}
\begin{abstract}
Group activities are important behaviors in human society, providing personalized recommendations for groups is referred to as the group recommendation task. Existing methods can usually be categorized into two strategies to infer group preferences: 1) determining group preferences by aggregating members' personalized preferences, and 2) inferring group consensus by capturing group members' coherent decisions after common compromises. However, the former would suffer from the lack of group-level considerations, and the latter overlooks the fine-grained preferences of individual users. To this end, we propose a novel group recommendation method \textbf{AlignGroup}, which focuses on both group consensus and individual preferences of group members to infer the group decision-making. Specifically, AlignGroup explores group consensus through a well-designed hypergraph neural network that efficiently learns intra- and inter-group relationships. Moreover, AlignGroup innovatively utilizes a self-supervised alignment task to capture fine-grained group decision-making by aligning the group consensus with members' common preferences. Extensive experiments on two real-world datasets validate that our AlignGroup outperforms the state-of-the-art on both the group recommendation task and the user recommendation task, as well as outperforms the efficiency of most baselines. 

\end{abstract}




\maketitle

\section{Introduction}
As social media becomes ubiquitous \cite{yang2022personality,wei2023lightgt}, the online platform has become the main position and key information source of group activities, such as Mafengwo facilitating travel planning and Meetup hosting social gatherings \cite{yin2020overcoming}. Traditional recommendation systems, designed for individuals, struggle to cope with the complexities of group decision-making. This gap has spurred the need for group recommendation systems, which aim to balance members' individual preferences and group consensus to deliver fine-grained recommendations. In the e-commerce domain, such systems are crucial, as seen from Facebook Events to Yelp, where they must respect individual tastes while capturing group desires to ensure the accuracy of recommendations \cite{chen2020try,zhang2021double}.

Echoing the methods of conventional recommender systems, latent factor models are pivotal in the domain of group recommendations. Specifically, groups and items are represented as vectors by mapping into the same semantic space. Then estimating the suitable items closes with the group decision-making. Recent works \cite{amer2009group,baltrunas2010group,boratto2011state,cao2018attentive,cao2019social,chen2022thinking,guo2020group,he2020game,vinh2019interact} on group recommendations have largely concentrated on aggregating the preferences of members to predict the decision-making of a group. Classic aggregation techniques rely on pre-defined heuristic rules, such as averaging \cite{boratto2011state}, minimizing misery \cite{amer2009group}, or maximizing satisfaction \cite{baltrunas2010group}. More recently, there are also advanced deep-learning models \cite{cao2019social,cao2018attentive,guo2020group,he2020game,vinh2019interact} that employ attention mechanisms to facilitate preference aggregation. For example, AGREE \cite{cao2018attentive} proposes an attentive group recommendation that dynamically aggregates the preferences of users within a group, and HyperGroup \cite{guo2021hierarchical} explores the superior representational capabilities of hypergraphs for group recommendation. CubeRec \cite{chen2022thinking} leverages the geometrically rich structure of hypercubes \cite{ren2020query2box,zhang2021learning} to integrate multifaceted member preferences. It is worth mentioning that ConsRec \cite{wu2023consrec} realizes that group consensus cannot be integrated by group users' preferences alone; it reveals the consensus hidden behind group behavior, achieving state-of-the-art performance.

\begin{figure}[t]
    \centering
    \subfigure[HR@5] {
        \label{fig:HR@5}
        \includegraphics[width=0.45\linewidth]{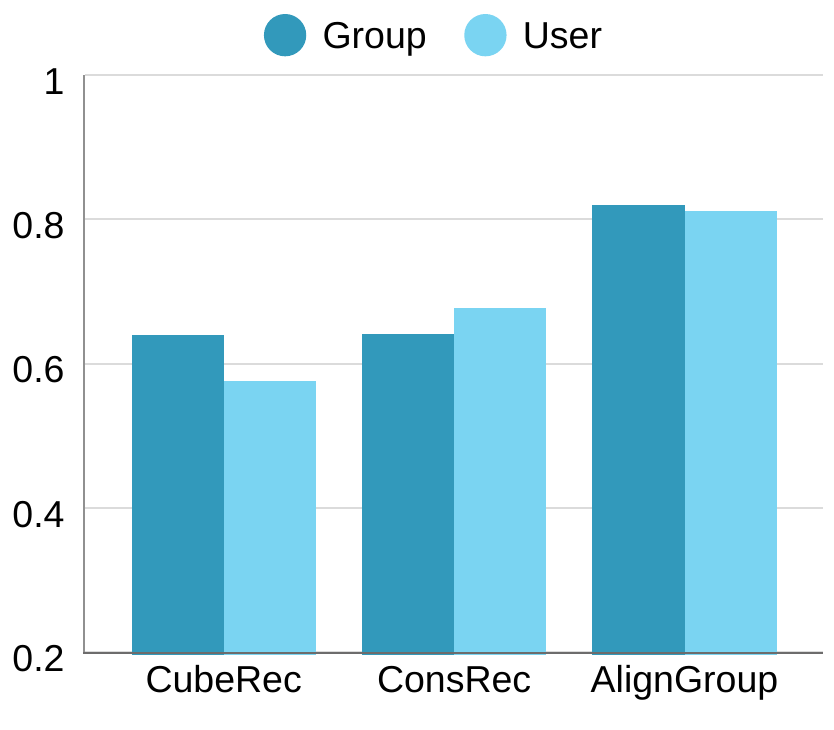}
        }  
    \subfigure[NDCG@5] {
        \label{fig:NDCG@5}
        \includegraphics[width=0.45\linewidth]{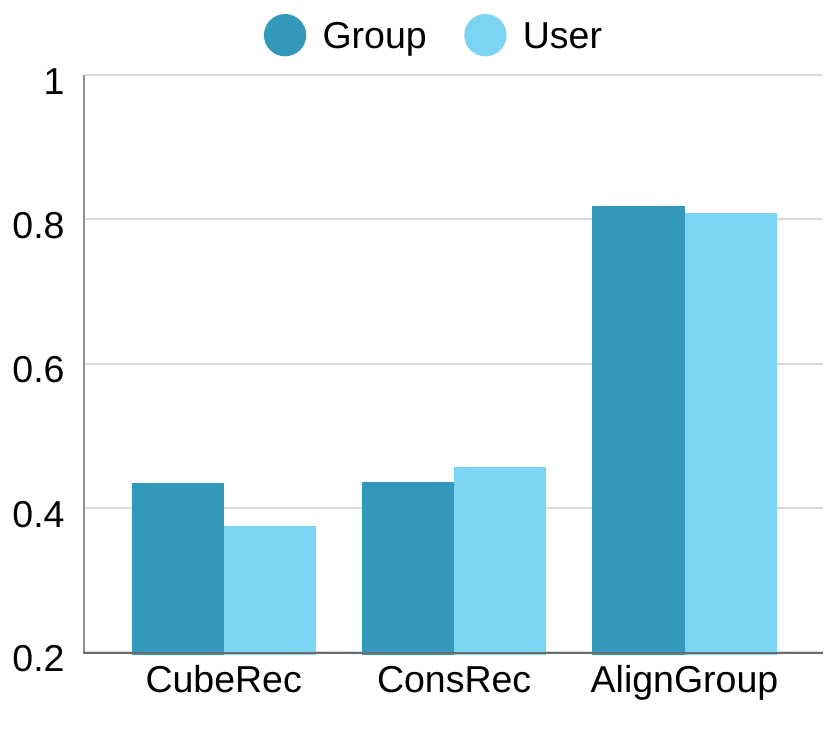}
        }
           \vskip -0.15in
    \caption{Performance comparison of CubeRec, ConsRec, and our AlignGroup on both group recommendation and user recommendation tasks in terms of Hit Ratio (HR) and Normalized Discounted Cumulative Gain (NDCG) on CAMRa2011.}   
    \label{fig:demo}
    \vskip -0.15in
\end{figure}

Nevertheless, the majority of approaches lack the capability to adequately capture group consensus, instead simply inferring the group decision-making by aggregating individual user preferences. Group consensus is not simply the aggregation of member preferences. For example, if a family of three goes to the movies, the kid likes animated movies, the parents like action movies, and ultimately the group consensus drives them to choose an educational movie. Although CubeRec adopts a geometric perspective, harnessing hypercubes in the semantic space to represent group consensus, the substantial volume of these hypercubes may compromise the accuracy of consensus representation. ConsRec extracts group consensus through multi-views of information. However, it greatly smoothes out the individual preferences of group members. Fig~\ref{fig:demo} displays the performance comparison of CubeRec, ConsRec, and our AlignGroup on the CAMRa2011 dataset for both group and user recommendation tasks. It is observed that CubeRec and ConsRec perform poorly on the Normalized Discounted Cumulative Gain (NDCG) compared with the Hit Ratio (HR). It is worth noting that the NDCG additionally considers the position of the demand item in the model's recommendation list compared to HR. Therefore, we point out that CubeRec and ConsRec are too coarse in inferring the group decision-making, and we further attribute this to the fact that both CubeRec and ConsRec only consider the group consensus but ignore individualized preferences for users within a group.

To this end, we propose a novel group recommendation method \textbf{AlignGroup}, which focuses on both group consensus and individual preferences of group members to infer the group decision-making. Specifically, we innovatively propose a hypergraph neural network to capture group consensus from both intra- and inter-group relations. The intra-group relations are learned by aggregating group member preferences and enhanced from high-order related neighbor groups' information through hypergraph convolution operations. For inter-group relations, we effectively capture the group relations from a member overlap perspective. Besides, to effectively infer user preferences and item properties in the same semantic space, we augment user and item representations by aggregating the related group consensus for each user or item. Furthermore, we introduce a novel and simple self-supervised alignment task to effectively capture the fine-grained group decision-making by aligning the group consensus with members' common preferences. In Section~\ref{sec:ssl-math}, we discuss different strategies and scopes for calculating members' common preferences, and further evaluate them in Section~\ref{sec:ssl}. 

In a nutshell, the contributions of our work are as follows:
\begin{itemize}[leftmargin=*]
    \item We propose a novel framework AlignGroup for the group recommendation, which can effectively infer the fine-grained group decision-making by considering both group consensus and members' preferences.
    \item We develop a new hypergraph neural network, which well-extracts group consensus from both intra- and inter-group relations.
    \item We design a self-supervised alignment task, effectively reducing the gap between member preferences and group consensus.
    \item We perform comprehensive experiments on two real-world datasets to validate both the effectiveness and efficiency of AlignGroup.
\end{itemize}

\section{Related work}
In this section, we review three related lines of research that contribute to group recommendation, namely methods for preference aggregation, hypergraph learning, and self-supervised learning.

\subsection{Preference Aggregation}
Earlier works use pre-defined heuristic rules to aggregate the preferences of users within a group to infer the group decision-making, such as averaging \cite{boratto2011state}, minimizing misery \cite{amer2009group}, or maximizing satisfaction \cite{baltrunas2010group}. However, in the real world, each member has a different status and role within the group. Therefore, pre-defined aggregation rules cannot dynamically adjust the weights. Recently, many deep learning-based aggregation methods have been proposed, which dynamically adjust the weight share of each user in a group through a learnable weight matrix. For example, MoSAN \cite{vinh2019interact} first utilizes the attention mechanism to capture fine-grained interactions among group members dynamically. Nevertheless, it only considers users as group members. AGREE \cite{cao2018attentive} broadens the scope of group members based on MoSAN, which considers both users and items as group members, and SoAGREE \cite{cao2019social} further exploits the relationships among users to enhance the user representation in AGREE. SIGR \cite{yin2019social} exploits and integrates the global and local social influence of users. CAGR \cite{yin2020overcoming} first tries to learn centrality-aware user representations and then leverages it to refer to group decision-making. GroupIM \cite{sankar2020groupim} adopts mutual information maximization between users and groups to overcome the data sparsity problem between items and groups. HyperGroup \cite{guo2021hierarchical} proposes a hyperedge embedding-based method to enhance the representations. It is worth noting that some recent works break the constraint of aggregating information, such as CubeRec \cite{chen2022thinking} exploits the geometric representational power of hypercubes to frame the scope of group consensus, and ConsRec \cite{wu2023consrec} utilizes multilevel views to collectively expose group consensus to infer group decision-making. However, we note that relying solely on group consensus or aggregating the preferences of group members is insufficient. Therefore, only by taking into account both individual preferences and group consensus can group decision-making be effectively inferred.

\subsection{Hypergraph Learning}
The relationship between the edge and vertex in the hypergraph is naturally similar to the relationship between group and member. Recent works \cite{guo2021hierarchical,jia2021hypergraph,zhang2021double,wu2023consrec} utilize hypergraphs to improve group recommendation performance. HyperGroup \cite{guo2021hierarchical} utilizes the edges of the hypergraph to represent groups, effectively exploiting the hypergraph structure to propagate in-group user collaboration signals. HCR \cite{jia2021hypergraph} utilizes a hypergraph structure to simultaneously propagate collaboration signals between users and items. S$^2$-HHGR \cite{zhang2021double} further utilizes the hypergraph to capture both in-group and out-group user relationships while utilizing self-supervised signals to align user representations at different granularities. ConsRec \cite{wu2023consrec} propagates multi-view collaborative signals from group and member perspectives to achieve state-of-the-art performance.

\subsection{Self-supervised Learning}
Recent works point out that self-supervised learning can also be effective in mitigating the data sparsity problem on the group recommendation task, GroupIM \cite{sankar2020groupim} utilizes self-supervised signals to maximize the mutual information of users and groups to improve the representation, S$^2$-HHGR \cite{zhang2021double} utilizes self-supervised signals to efficiently align user representations at different granularities, and CubeRec \cite{chen2022thinking} proposes a novel self-supervised signal to capture the common preferences of users within a hypercube.

\section{Preliminary}
Formally, we use bold lowercase letters (e.g., $\mathbf{e}$) and bold capital letters (e.g., $\mathbf{E}$) to represent vectors and matrices, respectively. None-bold letters (e.g., $x$) and squiggle letters (e.g., $\mathcal{X}$) are used to denote scalars and sets, respectively.

\subsection{Task Definition}
Let $\mathcal{U}$ = \{$u_1,...,u_{|\mathcal{U}|}$\}, $\mathcal{I}$ = \{$i_1,...,i_{|\mathcal{I}|}$\}, and $\mathcal{G}$ = \{$g_1,...,g_{|\mathcal{G}|}$\} be the sets of users, items, and groups respectively. There are two types of observed interactions. We use $\mathbf{R^U} \in \mathbb{R}^{|\mathcal{U}| \times |\mathcal{I}|}$ to denote the user-item interactions between $\mathcal{U}$ and $\mathcal{I}$, where the element $\mathbf{R^U(n,m)}$ = 1 if user $u_n$ has interacted with item $i_m$ otherwise $\mathbf{R^U(n,m)}$ = 0. Similarly, we use $\mathbf{R^G} \in \mathbb{R}^{|\mathcal{G}| \times |\mathcal{I}|}$ to denote the group-item interactions between $\mathcal{G}$ and $\mathcal{I}$. The group $g \in \mathcal{G}$ consists of a set of user members $\mathcal{U}_{g}$ = \{$u_1,...,u_p,...,u_{|\mathcal{U}_{g}|}$\} where $u_p \in \mathcal{U}$. We denote the interacted item set of $g$ as $\mathcal{I}_{g}$ = \{$i_1,...,i_q,...,i_{|\mathcal{I}_{g}|}$\} where $i_q \in \mathcal{I}$. The task for group recommendation is defined as recommending items that target group $g$ may be interested in.   

\subsection{Hypergraph}
Hypergraph is a more complex and enriched topological structure. Each hyperedge can contain more than one vertex. Formally, we define the hypergraph as $G$ = ($\mathcal{V},\mathcal{E},\mathcal{H}$) where $\mathcal{V} = \mathcal{U} + \mathcal{I}$ is the vertex set, $\mathcal{E}$ is the hyperedge set, and $\mathcal{H} \in \mathbb{R}^{|\mathcal{V}| \times |\mathcal{E}|}$ depicts the connectivity of the hypergraph as $\mathcal{H}(v,e)$ = 1 if the hyperedge $e$ connects the vertex $v$ otherwise $\mathcal{H}(v,e)$ = 0. Furthermore, let $\mathcal{E}_v$ denote a set of related hyperedges that connect to the node $v$ (i.e., $\mathcal{E}_v$ = \{$e \in \mathcal{E}|\mathcal{H}(v,e)$ = 1\}) and $\mathcal{V}_e$ denote a set of nodes that connect to the hyperedge $e$ (i.e., $\mathcal{V}_e$ = \{$v \in \mathcal{V}|\mathcal{H}(v,e)$ = 1\}). In the group recommendation scenario, all members of a group together form a hyperedge, each member is a vertex, and each vertex can belong to more than one hyperedge. Therefore, we use $g \in \mathcal{E}$ to represent hyperedge instead of $e \in \mathcal{E}$ in some cases.

\begin{figure*}
    \centering
    \includegraphics[width=0.9\linewidth]{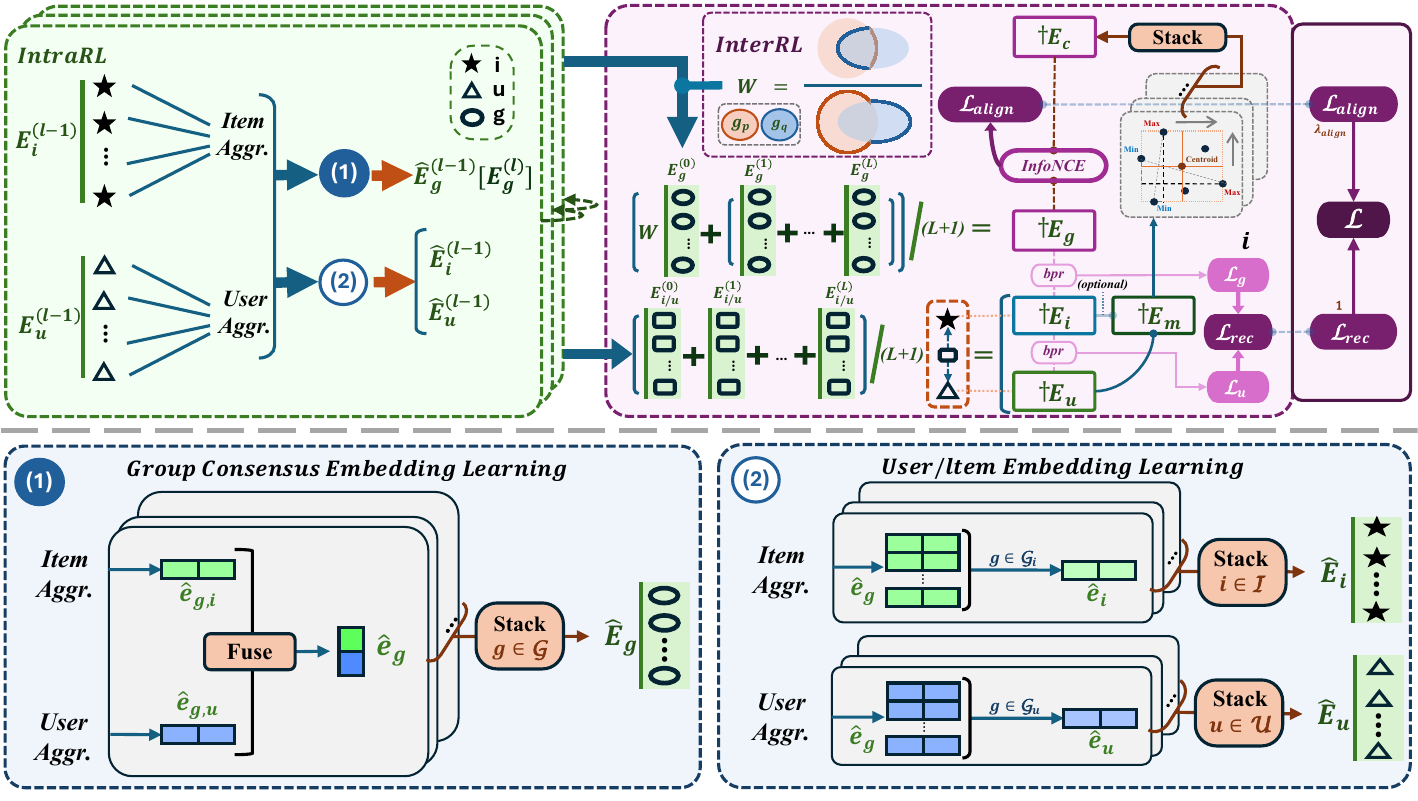}
    \vskip -0.15in
    \caption{AlignGroup overview. We construct a hypergraph neural network for both IntraRL and InterRL to capture group consensus. We further propose a self-supervised alignment task to align group consensus and members' common preferences. Steps (1) and (2) are the group consensus embedding learning and user/item embedding learning components within our hypergraph neural network.}
    \label{structure}
    \vskip -0.12in
\end{figure*}

\section{Methodology}
In this section, we elaborate on AlignGroup architecture and describe each component in detail. Specifically, AlignGroup first provides a hypergraph neural network to capture both group consensus embedding and member embedding (including both users and items) via both intra-group relation learning (IntraRL) and inter-group relation learning (InterRL). To ensure that the user's personalization preferences are not compromised too much for group consensus, we further introduce a self-supervised alignment signal to steer the group consensus closer to the users' preferences. Fig~\ref{structure} shows the overall architecture of AlignGroup.

\subsection{Hypergraph Neural Network}
In this subsection, we introduce a hypergraph neural network that aims to capture both group consensus and member preferences. Specifically, this hypergraph neural network aims to capture group consensus embedding $\mathbf{\dagger{E}}_g$, refined user embedding $\mathbf{\dagger{E}}_u$, and refined item embedding $\mathbf{\dagger{E}}_i$ via graph propagation. 

\subsubsection{Group Consensus Embedding Learning via Hypergraph Neural Network}
The group consensus is determined by the joint decision of all relevant members of the group, including users and items. At the same time, there is some similarity between groups with overlapping members.

We aim to extract both intra- and inter-group relations to effectively capture the group consensus. 

\textbf{Intra-group Relation Learning (IntraRL)}: First, we consider intra-group relation learning. We represent each group as a hyperedge $g \in \mathcal{E}$, including both users and items. However, user and item nodes preserve different semantic information. Therefore, we split the aggregation process of the hypergraph neural network into user-side and item-side. We feed the initial user embedding $\mathbf{E}_u \in \mathbb{R}^{|\mathcal{U}| \times d}$, initial item embedding $\mathbf{E}_i \in \mathbb{R}^{|\mathcal{I}| \times d}$, and initial group embedding $\mathbf{E}_g \in \mathbb{R}^{|\mathcal{G}| \times d}$ to the hypergraph neural network.

Formally, for hyperedge $g$, we compute the message for the user/item side within group $g$ via aggregation process:
\vskip -0.1in
\begin{equation}
\label{eq:1}
    \mathbf{e}_{u_{p}} = \mathbf{E}_u(p,:), \quad \mathbf{e}_{i_{q}} = \mathbf{E}_i(q,:),
\end{equation}
\begin{equation}
\label{eq:2}
\mathbf{\hat{e}}_{g,u} = \operatorname{Aggr}({\mathbf{e}_{u_{p}}|u_{p} \in \mathcal{U}_g}) = \sum_{u_{p} \in \mathcal{U}_g} \alpha_u \mathbf{e}_{u_{p}}, 
\end{equation}
\vskip -0.05in
\begin{equation}
\label{eq:3}
\mathbf{\hat{e}}_{g,i} = \operatorname{Aggr}({\mathbf{e}_{i_{q}}|i_{q} \in \mathcal{I}_g}) = \sum_{i_{q} \in \mathcal{I}_g} \alpha_i \mathbf{e}_{i_{q}}, 
\end{equation}
where $\alpha_i$ and $\alpha_u$ are the simple average operation (i.e., $\alpha_u = 1/{|\mathcal{U}_g|}$ and $\alpha_i = 1/{|\mathcal{I}_g|}$). Then we fuse the user-side message and the item-side message through a linear transformation:
\begin{equation}
\label{eq:4}
\mathbf{\hat{e}}_{g} = \operatorname{Concat}(\mathbf{\hat{e}}_{g,u}|\mathbf{\hat{e}}_{g,i})W,
\end{equation}
where $W \in \mathbb{R}^{2d \times d}$ denotes the trainable weight matrix for message fusion and $\operatorname{Concat}$ is the concatenation operation. Then we stack all $\mathbf{\hat{e}}_g (g \in \mathcal{G})$ to construct group consensus embeddings $\mathbf{\hat{E}}_g$.

Many works \cite{he2020lightgcn,xu2024mentor,yu2023xsimgcl,jiang2023adaptive} verify that multiple-layer Graph Convolution Networks (GCNs) can improve the expressiveness of representation. Both user/item representation and group consensus representation can benefit from high-order neighbors. To this end, we average the embedding obtained at each layer to compute the final group representation $\mathbf{\dagger{E}}_g$: 
\vskip -0.18in
\begin{equation}
\label{eq:5}
\mathbf{E}_g^{(l+1)} = \mathbf{\hat{E}}_g^{(l)}, \quad
\mathbf{\dagger{E}}_g= \frac{1}{L+1} \sum_{l=0}^{L} \mathbf{E}_g^{(l)},
\end{equation}
\vskip -0.05in
\noindent where $L$ is the total number of convolutional layers, $\mathbf{\hat{E}}_g$ is the next layer representation of $\mathbf{E}_g$.

We fully explore intra-group relations through the above steps. However, inter-group relations also deserve further consideration.

\begin{figure}
    \centering
    \includegraphics[width=0.99\linewidth]{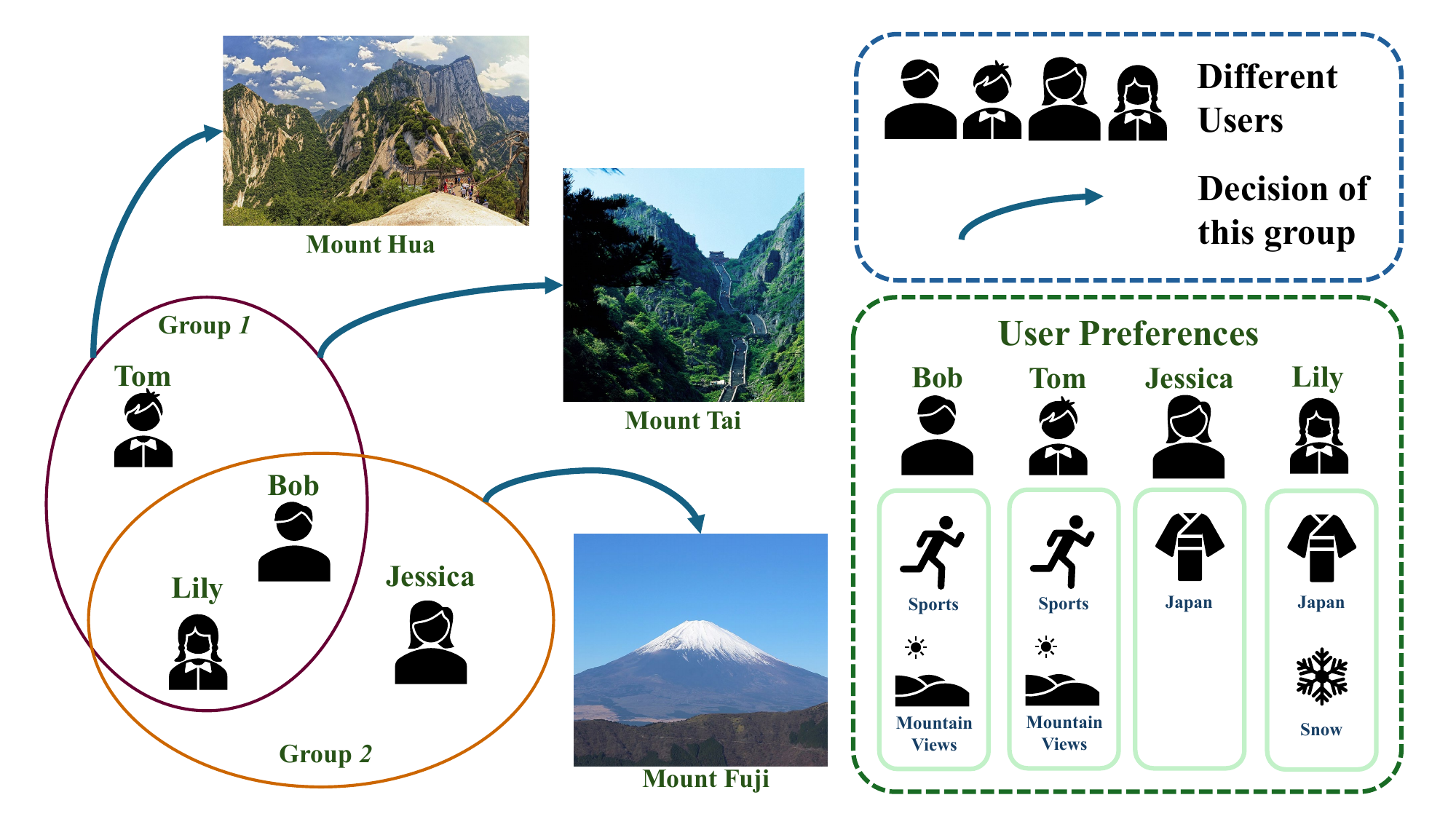}
    \vskip -0.15in
    \caption{An illustrative example of the inter-group relations.}
    \label{fig:2}
    \vskip -0.15in
\end{figure}

\begin{figure}
    \centering
    \includegraphics[width=0.99\linewidth]{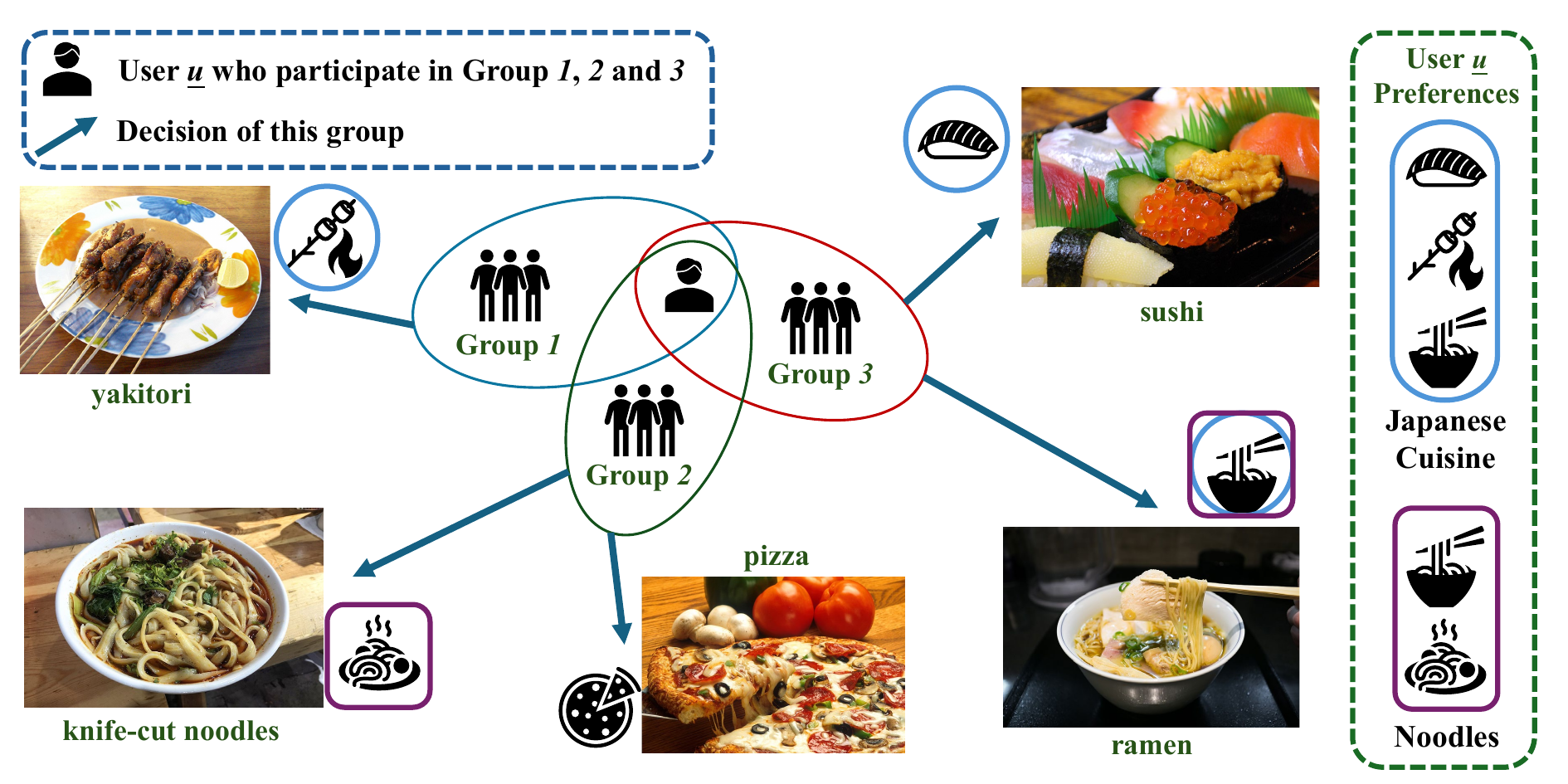}
    \vskip -0.15in
    \caption{An illustrative example of the user preferences.}
    \label{fig:3}
    \vskip -0.15in
\end{figure}

\textbf{Inter-group Relation Learning (InterRL)}: Inspired by ConsRec \cite{wu2023consrec}, we noticed that groups may have their inter-group relations. For example, as depicted in Fig~\ref{fig:2}, since Bob and Tom both like sports and mountain views, group 1 always chooses to climb the steeper Mount Hua and Mount Tai. However, in group 2, Bob is the only one who likes sports, and both Jessica and Lily like Japan, so they finally choose to enjoy Mount Fuji in Japan, which has a beautiful mountain view. It means that the degree of member overlap plays a significant role in the similarity between groups. To capture the group inherent preference among similar groups, we introduce a group overlap graph to capture the similarity between groups from a member overlap perspective.

Formally, we build an adjacency matrix $W_g$ to discriminate relevance between groups. $W_g(g_p,g_q)$ means the overlap ratio between group $g_p$ and group $g_q$:
\vskip -0.1in
\begin{equation}
\label{eq:6}
W_g(g_p,g_q) = \frac{|\mathcal{V}_{g_p} \cap \mathcal{V}_{g_q}|}{|\mathcal{V}_{g_p} \cup \mathcal{V}_{g_q}|}.
\end{equation}
\vskip -0.05in
We further combine inter- and intra-group relations. Specifically, we utilize a group overlap graph to explore the inter-group relations together with intra-group consensus embeddings $\mathbf{E}_g$, formally:
\begin{equation}
\label{eq:7}
\mathbf{\dagger{E}}_g = \frac{1}{L+1}(\sum_{l=1}^{L} \mathbf{E}_g^{(l)} + \mathbf{E}_g^{(0)}W_g).
\end{equation}
The first layer of GCN plays the most important role, so we effectively combine inter- and intra-group relations in the first layer of GCN. We further verify the importance of InterRL through an ablation study in Section~\ref{sec:InterRL}.

\subsubsection{User/Item Embedding Learning via Hypergraph Neural Network}
Intuitively, the preferences of a user or the properties of an item can be jointly modeled by the behavior of all groups to which it belongs. For example, as described in Fig~\ref{fig:3}, the preferences of user $u$ can be modeled by group consensus. Specifically, we can infer the preference of user $u$ for Japanese cuisine by observing their observed liking for yakitori in Group 1 and for ramen and sushi in Group 3. Additionally, the preference of user $u$ for noodles can be deduced from their fondness for knife-cut noodles in Group 2 and ramen in Group 3. It is worth noting that the pizza in Group 2 may mean user $u$ chooses to compromise with the group consensus.

To enhance the representation of user/item preference, we aggregate all related representations of each user/item by collecting the message from related hyperedges:
\begin{equation}
\label{eq:8}
\mathbf{\hat{e}}_u = \operatorname{Aggr}(\mathbf{\hat{e}}_{g}|g \in \mathcal{E}_u) = \sum_{g \in \mathcal{E}_u} \frac{1}{|\mathcal{E}_u|} \frac{\sum_{g' \in \mathcal{E}_u} (|\mathcal{U}_{g'}| + |\mathcal{I}_{g'}|)}{|\mathcal{E}_u|(|\mathcal{U}_g| + |\mathcal{I}_g|)} \mathbf{\hat{e}}_{g},
\end{equation}
\begin{equation}
\label{eq:9}
\mathbf{\hat{e}}_i = \operatorname{Aggr}(\mathbf{\hat{e}}_{g}|g \in \mathcal{E}_i) = \sum_{g \in \mathcal{E}_i} \frac{1}{|\mathcal{E}_i|} \frac{\sum_{g' \in \mathcal{E}_i} (|\mathcal{U}_{g'}| + |\mathcal{I}_{g'}|)}{|\mathcal{E}_i|(|\mathcal{U}_g| + |\mathcal{I}_g|)} \mathbf{\hat{e}}_{g},
\end{equation}
where $\mathbf{\hat{e}}_u$ and $\mathbf{\hat{e}}_i$ means the enhanced embedding of user $u$ and item $i$, respectively. In the real world, the more people in a group, the lower the correlation between group consensus and individual preferences. Therefore, we utilize group size to dynamically adjust the influence weights of different group consensus on individual preferences. $\sum_{g' \in \mathcal{E}_u} (|\mathcal{U}_{g'}| + |\mathcal{I}_{g'}|)/|\mathcal{E}_u|(|\mathcal{U}_g| + |\mathcal{I}_g|)$ and $\sum_{g' \in \mathcal{E}_i} (|\mathcal{U}_{g'}| + |\mathcal{I}_{g'}|)/|\mathcal{E}_i|(|\mathcal{U}_g| + |\mathcal{I}_g|)$ are the adjust weights. Then we stack all $\mathbf{e}_u (u \in \mathcal{U})$ and $\mathbf{e}_i (i \in \mathcal{I})$ to construct user and item embeddings $\mathbf{\hat{E}}_u$ and $\mathbf{\hat{E}}_i$, respectively.

We further enhance the expressiveness of $\mathbf{\hat{E}}_u$ and $\mathbf{\hat{E}}_i$ by GCNs. Specifically, we average the embedding obtained at each layer to get the final user representation $\mathbf{\dagger{E}}_u$ and item representation $\mathbf{\dagger{E}}_i$:
\vskip -0.15in
\begin{equation}
\label{eq:10}
\mathbf{E}_u^{(l+1)} = \mathbf{\hat{E}}_u^{(l)}, \quad
\mathbf{\dagger{E}}_u = \frac{1}{L+1} \sum_{l=0}^{L} \mathbf{E}_u^{(l)},
\end{equation}
\vskip -0.1in
\begin{equation}
\label{eq:11}
\mathbf{E}_i^{(l+1)} = \mathbf{\hat{E}}_i^{(l)}, \quad
\mathbf{\dagger{E}}_i = \frac{1}{L+1} \sum_{l=0}^{L} \mathbf{E}_i^{(l)},
\end{equation}
\vskip -0.05in
\noindent where $L$ is the total number of convolutional layers, $\mathbf{\hat{E}}_u$ and $\mathbf{\hat{E}}_i$ are the next layer representation of $\mathbf{E}_u$ and $\mathbf{E}_i$, respectively.

Algorithm~\ref{hnn} shows the process of our hypergraph neural network. 

\begin{algorithm}
\caption{Hypergraph Neural Network}
\label{hnn}
\begin{algorithmic} [1] 
\STATE \textbf{Input:} $\mathcal{U}$, $\mathcal{I}$, $\mathcal{G}$, GCN Layer number $L$, $\mathcal{V}$, $\mathcal{E}$, $\mathcal{H}$
\STATE \textbf{Output:} User embedding $\mathbf{\dagger{E}}_u$, Item embedding $\mathbf{\dagger{E}}_i$, Group embedding $\mathbf{\dagger{E}}_g$
\STATE Initialize $\mathbf{E}_u$, $\mathbf{E}_i$, $\mathbf{E}_g$;
\STATE $\mathbf{E}_u^{(0)}$ $\gets$ $\mathbf{E}_u$; $\mathbf{E}_i^{(0)}$ $\gets$ $\mathbf{E}_i$; $\mathbf{E}_g^{(0)}$ $\gets$ $\mathbf{E}_g$;
\FOR{$l = 1...L$}
    \FOR{$g \in \mathcal{G}$} 
        \STATE $\mathbf{\hat{e}}_{g,u}$ $\gets$ $\operatorname{Aggr}(\mathbf{e}_{u_p}|u_p \in \mathcal{U}_g)$ with Eq.\ref{eq:2};
        \STATE $\mathbf{\hat{e}}_{g,i}$ $\gets$ $\operatorname{Aggr}(\mathbf{e}_{i_q}|i_q \in \mathcal{I}_g)$ with Eq.\ref{eq:3};
        \STATE $\mathbf{\hat{e}}_{g}$ $\gets$ $\operatorname{Concat(\mathbf{\hat{e}}_{g,u},\mathbf{\hat{e}}_{g,i}})W$;
    \ENDFOR
    \STATE $\mathbf{\hat{E}}_g$ $\gets$ $\operatorname{Stack}(\mathbf{\hat{e}}_{g}|g \in \mathcal{G}$);
    \FOR{$u \in \mathcal{U}$} 
        \STATE $\mathbf{\hat{e}}_{u}$ $\gets$ $\operatorname{Aggr}(\mathbf{\hat{e}}_{g}|g \in \mathcal{E}_u)$ with Eq.\ref{eq:8};
    \ENDFOR
    \STATE $\mathbf{\hat{E}}_u$ $\gets$ $\operatorname{Stack}(\mathbf{\hat{e}}_{u}|u \in \mathcal{U}$);
    \FOR{$i \in \mathcal{I}$} 
        \STATE $\mathbf{\hat{e}}_{i}$ $\gets$ $\operatorname{Aggr}(\mathbf{\hat{e}}_{g}|g \in \mathcal{E}_i)$ with Eq.\ref{eq:9};
    \ENDFOR
    \STATE $\mathbf{\hat{E}}_i$ $\gets$ $\operatorname{Stack}(\mathbf{\hat{e}}_{i}|i \in \mathcal{I}$);
    \STATE $\mathbf{E}_g^{(l)}$ $\gets$ $\mathbf{\hat{E}}_g$; $\mathbf{E}_u^{(l)}$ $\gets$ $\mathbf{\hat{E}}_u$; $\mathbf{E}_i^{(l)}$ $\gets$ $\mathbf{\hat{E}}_i$;
\ENDFOR
\FOR{$(g_p,g_q) \in \mathcal{G}$} 
    \STATE $W_g(g_p,g_q)$ $\gets$ $\frac{|\mathcal{V}_{g_p} \cap \mathcal{V}_{g_q}|}{|\mathcal{V}_{g_p} \cup \mathcal{V}_{g_q}|}$;
\ENDFOR
\STATE $\mathbf{\dagger{E}}_g$ $\gets$ $\frac{1}{L+1}(\sum_{l=1}^{L} \mathbf{E}_g^{(l)} + \mathbf{E}_g^{(0)}W_g)$;
\STATE $\mathbf{\dagger{E}}_u$ $\gets$ $\frac{1}{L+1}(\sum_{l=0}^{L} \mathbf{E}_u^{(l)})$; $\mathbf{\dagger{E}}_i$ $\gets$ $\frac{1}{L+1}(\sum_{l=0}^{L} \mathbf{E}_i^{(l)})$;
\end{algorithmic}
\end{algorithm}
\vskip -0.1in

\subsection{Self-supervised Alignment Task}
\label{sec:ssl-math}
Unfortunately, in many cases, members have to compromise their individual preferences to reach the group consensus. To satisfy both group consensus and member preferences, we propose a self-supervised alignment task to align group consensus and members' common preferences.

Two important problems need to be identified, \textbf{a) the scope of group members}, and \textbf{b) how to calculate the common preferences of members within a group}.

We present our solution to achieve effective alignment between group consensus and member preferences. Specifically, we introduce two different scopes of group members and two different strategies for calculating the common preferences of members within a group. We further testify these various scopes and strategies via ablation study in Section~\ref{sec:ssl}.

\subsubsection{The Scope of Group Members} In the hypergraph neural network section, we treat users and items as members of groups equally. However, we cannot deny that user preferences can be compromised for group consensus, while properties of items are fixed properties. Therefore, we propose two scopes in computing common member preferences: \textbf{only users}, \textbf{both users and items}.

\subsubsection{How to Calculate the Common Preferences of Members within a Group} For a given group $g \in \mathcal{G}$, it consists of a set of members $\mathcal{M}_{g}$ = $\{m_1,...m_k,...,m_{|\mathcal{M}_{g}|}\}$, where $m_k \in \mathcal{U}_g$ or $m_k \in \mathcal{U}_g \cup \mathcal{I}_g$. 

Thanks to the hypergraph neural network part, we have projected the embedding of item, user, and group into the same semantic space. Therefore, we consider defining the common preferences of group members from a physical space perspective.

\textbf{Geometric Centroid:}
The geometric centroid for group $g$ can be computed as the following:
\begin{equation}
\label{eq:12}
    \mathbf{e}_{g,c} = \frac{\operatorname{Max} (\mathbf{\dagger{e}}_{m_k}|m_k \in \mathcal{M}_{g}) + \operatorname{Min} (\mathbf{\dagger{e}}_{m_k}|m_k \in \mathcal{M}_{g})}{2},
\end{equation}
where $\operatorname{Max}$ and $\operatorname{Min}$ both operate element-wise. $\mathbf{\dagger{e}}_{m_k}$ is the vector of $\mathbf{\dagger{E}}_{m_k} \in \mathbf{\dagger{E}}_{u} \cup \mathbf{\dagger{E}}_{i}$.

\textbf{Geometric Barycenter:}
The geometric barycenter for group $g$ can be computed as the following:
\begin{equation}
\label{eq:13}
    \mathbf{e}_{g,c} = \operatorname{Avg}(\mathbf{\dagger{e}}_{m_k}|m_k \in \mathcal{M}_{g}),
\end{equation}
where $\operatorname{Avg}$ operate element-wise. $\mathbf{\dagger{e}}_{m_k}$ is the vector of $\mathbf{\dagger{E}}_{m_k}$.

Then we stack all $\mathbf{e}_{g,c} (g \in \mathcal{G})$ to construct members' common preferences embeddings $\mathbf{\dagger{E}}_c$.

\subsubsection{Loss Function}
We adopt InfoNCE \cite{oord2018representation} to align group consensus embeddings $\mathbf{\dagger{E}}_g$ and members' common preferences embeddings $\mathbf{\dagger{E}}_c$. Formally, the alignment learning loss is defined as:
\vskip -0.15in
\begin{equation}
\label{eq:14}
\mathcal{L}_{align} =\sum_{g_p \in \mathcal{G}}-\log \frac{\exp \Big(\mathbf{\dagger{e}}_{g_p,c} \cdot \mathbf{\dagger{e}}_{g_p} / \tau\Big)}{\sum_{g_q \in \mathcal{G}} \exp \Big(\mathbf{\dagger{e}}_{g_q,c} \cdot \mathbf{\dagger{e}}_{g_q} / \tau\Big)},
\end{equation}
where $\mathbf{\dagger{e}}_{g_p,c}$ and $\mathbf{\dagger{e}}_{g_q,c}$ are the vectors of $\mathbf{\dagger{E}}_{c}$, and $\mathbf{\dagger{e}}_{g_p}$ and $\mathbf{\dagger{e}}_{g_q}$ are the vectors of $\mathbf{\dagger{E}}_{g}$. $\tau$ is the temperature hyper-parameter. 

Overview steps can be found in Algorithm \ref{ssa}.

\vskip -0.1in
\begin{algorithm}
\caption{Self-supervised Alignment Task}
\label{ssa}
\begin{algorithmic} [1] 
\STATE \textbf{Input:} $\mathcal{U}$, $\mathcal{I}$, $\mathcal{G}$, $\mathbf{\dagger{E}}_{u}$, $\mathbf{\dagger{E}}_{i}$, $\mathbf{\dagger{E}}_{g}$
\STATE \textbf{Output:} Self-supervised alignment loss $\mathcal{L}_{align}$ 
\FOR{$g \in \mathcal{G}$}
    \STATE $\mathcal{M}_g$ $\gets$ $\mathcal{U}_g$ or $\mathcal{M}_g$ $\gets$ $\mathcal{U}_g \cap \mathcal{I}_g$;
    \IF{\textbf{Geometric Centroid}}
        \STATE $\mathbf{e}_{g,c}$ $\gets$ Compute with Eq~\ref{eq:12} \textbf{\# Geometric Centroid};
    \ELSE{}
        \STATE $\mathbf{e}_{g,c}$ $\gets$ Compute with Eq~\ref{eq:13} \textbf{\# Geometric Barycenter};
    \ENDIF
\ENDFOR
\STATE $\mathbf{\dagger{E}}_c$ $\gets$ $\operatorname{Stack}(\mathbf{e}_{g,c}|g \in \mathcal{G}$);
\STATE $\mathcal{L}_{align}$ $\gets$ $\operatorname{InfoNCE}(\mathbf{\dagger{E}}_g, \mathbf{\dagger{E}}_c)$ with Eq~\ref{eq:14};
\end{algorithmic}
\end{algorithm}
\vskip -0.1in

\subsection{Optimization}
We introduce our optimization strategy that jointly learns user-item and group-item interactions. Specifically, we provide a shared Multi-layer Perceptron (MLP) as a prediction function to compute both user-item final scores and group-item final scores. Formally:
\begin{equation}
\label{eq:15}
\operatorname{MLP}(\mathbf{e}) = \operatorname{LeakyReLU}(\mathbf{e}W_1)W_2,
\end{equation}
\begin{equation}
\label{eq:16}
\hat{y}_{u,i} = \sigma(\operatorname{MLP}(\mathbf{\dagger{e}}_u \odot \mathbf{\dagger{e}}_i)), 
\quad
\hat{y}_{g,i} = \sigma(\operatorname{MLP}(\mathbf{\dagger{e}}_g \odot \mathbf{\dagger{e}}_i)), 
\end{equation}
where $W_1 \in \mathbb{R}^{d \times 8}$ and $W_2 \in \mathbb{R}^{8 \times 1}$ denote the trainable weight matrices, $\mathbf{\dagger{e}}_u$, $\mathbf{\dagger{e}}_i$, and $\mathbf{\dagger{e}}_g$ are the vectors of $\mathbf{\dagger{E}}_u$, $\mathbf{\dagger{E}}_i$, and $\mathbf{\dagger{E}}_g$, respectively. $\sigma$ denotes the Sigmoid function, and $\odot$ is element-wise multiplication. We choose $\operatorname{LeakyReLU}$ \cite{maas2013rectifier} to solve the neuron death problem for ReLU.

With both the user-item interaction and group-item interaction data, we utilize the Bayesian Personalized Ranking (BPR) \cite{rendle2012bpr} loss for optimization:
\begin{equation}
\label{eq:17}
\mathcal{L}_{u}=-\sum_{u \in \mathcal{U}} \frac{1}{\left|\mathcal{D}_{u}\right|} \sum_{\left(p, n\right) \in \mathcal{D}_{u}} \hat{y}_{u,p}-\hat{y}_{u,n},
\end{equation}
\begin{equation}
\label{eq:18}
\mathcal{L}_{g}=-\sum_{g \in \mathcal{G}} \frac{1}{\left|\mathcal{D}_{g}\right|} \sum_{\left(p, n\right) \in \mathcal{D}_{g}} \hat{y}_{g,p}-\hat{y}_{g,n},
\end{equation}
where $\mathcal{D}_{u}$ and $\mathcal{D}_{g}$ denote the user-item training set sampled for user $u$ and group-item training set sampled for group $g$, respectively. For each ($p,n$) pair, $p$ and $n$ denote the observed item and unobserved item, respectively.

We jointly train $\mathcal{L}_{u}$ and $\mathcal{L}_{g}$ on all the user-item and group-item interactions as:
\begin{equation}
\label{eq:19}
\mathcal{L}_{rec} = \mathcal{L}_{u} + \mathcal{L}_{g}.
\end{equation}
Finally, we combine the $\mathcal{L}_{rec}$ with auxiliary self-supervised alignment task $\mathcal{L}_{align}$ to jointly training:
\begin{equation}
\label{eq:20}
\mathcal{L} = \mathcal{L}_{rec} + \lambda_{align}\mathcal{L}_{align},
\end{equation}
where $\lambda_{align}$ is the balancing hyper-parameter.

\section{Experiment}
In this section, we conduct a comprehensive experiment to validate the effectiveness of our proposed AlignGroup. Specifically, we aim to answer the following research questions (RQs): 

\begin{itemize}[leftmargin=*]
    \item RQ1: Does AlignGroup outperform the state-of-the-art group recommendation methods?
    \item RQ2: Whether the InterRL is necessary for AlignGroup?
    \item RQ3: How do different strategies and scopes of our self-supervised alignment task affect the performance?
    \item RQ4: Can AlignGroup effectively align group consensus and member preferences?
    \item RQ5: How efficient is AlignGroup compared with other group recommendation methods?
    \item RQ6: How do different hyper-parameter settings impact the performance of AlignGroup?
\end{itemize}

\begin{table}[!ht]
    \centering
    \vskip -0.1in
\caption{Statistics of the experimental datasets.}
\label{tab:dataset_statistics}
\vskip -0.15in
\resizebox{\linewidth}{!}{
    \begin{tabular}{c|ccccccc}
    \hline
         Dataset&  \# Users&  \# Items&  \# Groups& \makecell[c]{\# U-I \\interactions}&\makecell[c]{\# G-I \\interactions} & \makecell[c]{Avg. \\ group size}\\
         \hline
         Mafengwo & 5275 & 1513 & 995 & 39761 & 3595 & 7.19\\
         CAMRa2011 & 602 & 7710 & 290 & 116344 & 145068 & 2.08\\
         \hline
    \end{tabular}
    }
    \vskip -0.1in
\end{table}

\subsection{Experimental Settings}
\subsubsection{Datasets}
To evaluate our AlignGroup\footnote[1]{The code is available on \href{https://github.com/Jinfeng-Xu/AlignGroup}{https://github.com/Jinfeng-Xu/AlignGroup}.} in both group and user recommendation tasks, we conduct extensive experiments on two real-world public datasets, Mafengwo and CAMRa2011 \cite{cao2018attentive}. 

\textbf{1. Mafengwo: }The Mafengwo dataset encompasses travel records from a social tourism platform, detailing 5,275 users, 995 groups, and 1,513 locations with interactions between users and visited sites. This dataset was curated by selecting groups with at least two members who have visited three or more venues, averaging 7.19 users per group.

\textbf{2. CAMRa2011: }The CAMRa2011 dataset provides a collection of movie ratings from individual and household users, comprising 602 users in 290 groups with a total of 7,710 movies rated on a scale from 0 to 100. To focus on communal viewing preferences, only users belonging to groups were retained, with ratings converted into binary feedback, resulting in 116,344 user-item and 145,068 group-item interactions, with an average group size of 2.08.

The key statistics of these two datasets are delineated in Table~\ref{tab:dataset_statistics}.

\subsubsection{Evaluation Metrics}
To evaluate the recommendation task performance fairly, we adopt two well-established metrics following previous settings \cite{cao2018attentive,cao2019social}: Hit Ratio (HR) and Normalized Discounted Cumulative Gain (NDCG). We report the average metrics of all users in the test dataset under HR@5 (H@5), HR@10 (H@10), NDCG@5 (N@5), and NDCG@10 (N@10).

\begin{table*}[!t]
    \centering
\caption{Performance comparison of baselines and AlignGroup on group recommendation in terms of H@K and N@K. The superscript $^*$ indicates the improvement is statistically significant where the p-value is less than 0.01.}
\vskip -0.1in
\label{tab:group-performance}
\resizebox{\linewidth}{!}{
    \begin{tabular}{c|c|ccccccccc|cc}
    \hline
         Dataset & Metric & Pop & NCF & AGREE & HyperGroup & HCR & GroupIM & S$^2$-HHGR & CubeRec & ConsRec & AlignGroup & \textit{Improv.}\\
         \hline
         \multirow{4}{*}{Mafengwo}
          & H@5 & 0.3115 &0.4701& 0.4729 & 0.5739 & 0.7759 & 0.7377 & 0.7568 & 0.8613 & \underline{0.8844} & \textbf{0.9678$^*$} &  9.43\%\\
          & H@10 & 0.4251 & 0.6269 & 0.6321 & 0.6482 & 0.8503 & 0.8161 & 0.7779 & 0.9025 & \underline{0.9156} & \textbf{0.9688$^*$} &  5.81\%\\
          & N@5 & 0.2169 & 0.3657 & 0.3694 & 0.4777 & 0.6611 & 0.6078 & 0.7322 & 0.7574 & \underline{0.7692} & \textbf{0.8817$^*$} &  14.63\%\\
          & N@10 & 0.2537 & 0.4141 & 0.4203 & 0.5018 & 0.6852 & 0.6330 & 0.7391 & 0.7708 & \underline{0.7794} & \textbf{0.8821$^*$} &  13.18\%\\
          \hline
         \multirow{4}{*}{CAMRa2011}
          & H@5 & 0.4324 & 0.5803 & 0.5879 & 0.5890 & 0.5883 & \underline{0.6552} & 0.6062 & 0.6400 & 0.6407 & \textbf{0.8193$^*$} &  25.05\%\\
          & H@10 & 0.5793 & 0.7693 & 0.7789 & 0.7986 & 0.7821 & \underline{0.8407} & 0.7903 & 0.8207 & 0.8248 & \textbf{0.8448} &  0.49\%\\
          & N@5 & 0.2825 & 0.3896 & 0.3933 & 0.3856 & 0.4044 & 0.4310 & 0.3853 & 0.4346 & \underline{0.4358} & \textbf{0.8180$^*$} &  87.70\%\\
          & N@10 & 0.3302 & 0.4448 & 0.4530 & 0.4538 & 0.4670 & 0.4914 & 0.4453 & 0.4935 & \underline{0.4945} & \textbf{0.8259$^*$} &  67.02\%\\
         \hline
    \end{tabular}
    }
    \vskip -0.1in
\end{table*}

\begin{table*}[!t]
    \centering
\caption{Performance comparison of baselines and AlignGroup on user recommendation in terms of H@K and N@K. The superscript $^*$ indicates the improvement is statistically significant where the p-value is less than 0.01.}
\vskip -0.1in
\label{tab:user-performance}
\resizebox{\linewidth}{!}{
    \begin{tabular}{c|c|ccccccccc|cc}
    \hline
         Dataset & Metric & Pop & NCF & AGREE & HyperGroup & HCR & GroupIM & S$^2$-HHGR & CubeRec & ConsRec & AlignGroup & \textit{Improv.}\\
         \hline
         \multirow{4}{*}{Mafengwo}
          & H@5 & 0.4047 & 0.6363 & 0.6357 & 0.7235 & 0.7571 & 0.1608 & 0.6380 & 0.1847 & \underline{0.7725} & \textbf{0.8159$^*$} &  5.62\%\\
          & H@10 & 0.4971 & 0.7417 & 0.7403 & 0.7759 & 0.8290 & 0.2497 & 0.7520 & 0.3734 & \underline{0.8404} & \textbf{0.8643$^*$} &  2.84\%\\
          & N@5 & 0.2876 & 0.5432 & 0.5481 & 0.6722 & 0.6703 & 0.1134 & 0.4637 & 0.1099 & \underline{0.6884} & \textbf{0.7605$^*$} &  10.47\%\\
          & N@10 & 0.3172 & 0.5733 & 0.5738 & 0.6894 & 0.6937 & 0.1420 & 0.5006 & 0.1708 & \underline{0.7107} & \textbf{0.7763$^*$} &  9.23\%\\
          \hline
         \multirow{4}{*}{CAMRa2011}
          & H@5 & 0.4624 & 0.6119 & 0.6196 & 0.5728 & 0.6262 & 0.6113 & 0.6153 & 0.5754 & \underline{0.6774} & \textbf{0.8116$^*$} &  19.81\%\\
          & H@10 & 0.6026 & 0.7894 & 0.7897 & 0.7601 & 0.7924 & 0.7771 & 0.8173 & 0.7827 & \textbf{0.8412} & \underline{0.8382} &  -\\
          & N@5 & 0.3104 & 0.4018 & 0.4098 & 0.4410 & 0.4195 & 0.4064 & 0.3978 & 0.3751 & \underline{0.4568} & \textbf{0.8083$^*$} &  76.95\%\\
          & N@10 & 0.3560 & 0.4535 & 0.4627 & 0.5016 & 0.4734 & 0.4606 & 0.4641 & 0.4428 & \underline{0.5104} & \textbf{0.8167$^*$} &  60.01\%\\
         \hline
    \end{tabular}
    }
    \vskip -0.1in
\end{table*}

\subsubsection{Baselines}
We compare AlignGroup against the following state-of-the-art recommendation methods:
\begin{itemize}[leftmargin=*]
    \item \textbf{Popularity (Pop)} \cite{cremonesi2010performance} is a non-personalized method that ranks items for users and groups according to how often items are interacted with, providing a non-personalized standard for comparing personalized recommendation methods.
    \item \textbf{NCF} \cite{he2017neural} is adopted as treating each group as a virtual user for prediction.
    \item \textbf{AGREE} \cite{cao2018attentive} is a group recommendation method that uses an attention mechanism to aggregate user preferences within a group.
    \item \textbf{HyperGroup} \cite{guo2021hierarchical} treats a group as a hyperedge and introduces a method for learning group representations based on hyperedge embeddings.
    \item \textbf{HCR} \cite{jia2021hypergraph} is a notable group recommendation model that uses a dual-channel hypergraph convolutional network to analyze both member and group preferences.
    \item \textbf{GroupIM} \cite{sankar2020groupim} combines user preferences into group preferences using attention and includes self-supervised learning to address data sparsity by linking users with their groups.
    \item \textbf{S$^2$-HHGR} \cite{zhang2021double} is a state-of-the-art method that combines hypergraph and self-supervised learning for improved prediction of group preferences.
    \item \textbf{CubeRec} \cite{chen2022thinking} utilizes the geometric expressiveness of hypercubes to adaptively aggregate members’ interests.
    \item \textbf{ConsRec} \cite{wu2023consrec} reveals the group consensus by aggregate member-level, item-level, and group-level views.
\end{itemize}

\subsubsection{Implementation Details}
Following the basic settings of ConsRec \cite{wu2023consrec}, we implement AlignGroup in PyTorch and optimize with Adam optimizer. We apply Xavier initialization \cite{glorot2010understanding} for all initial random embedding. For hidden layers, we randomly initialize their parameters with a Gaussian distribution of a mean of 0 and a standard deviation of 0.1. As for hyper-parameter settings on AlignGroup, we perform a grid search on the number of convolutional layers $L$ in \{1, 2, 3, 4\}, temperature hyper-parameter $\tau$ in \{0.2, 0.4, 0.6, 0.8\}, and the balancing hyper-parameter $\lambda_{align}$ in \{1e-1, 1e-2, 1e-3\}. We fix the learning rate with 1e-3.

\subsection{Overall Performance (RQ1)}
Table~\ref{tab:group-performance} and Table~\ref{tab:user-performance} show the performance of our proposed AlignGroup and other baseline methods on the group recommendation and user recommendation tasks, respectively. We have the following key observations:
\begin{itemize}[leftmargin=*]
    \item For the group recommendation task shown in Table~\ref{tab:group-performance}, AlignGroup achieves higher performance than all baselines, which demonstrates the effectiveness of our AlignGroup. We attribute this to the excellent ability of our IntraRL and InterRL in hypergraph neural networks to capture group consensus. In addition, we note that the improvement of AlignGroup compared to other baselines is huge for NDCG, which is more concerned with fine-grained ranking order than HR. Therefore, we believe this implies that merely obtaining group consensus lacks fine-grained inference over group behavior, such as ConsRec and CubeRec. On the contrary, our self-supervised alignment task can effectively capture the fine-grained behavior of the group by aligning the group consensus with the members' common preferences.
    \item For the user recommendation task shown in Table~\ref{tab:user-performance}, AlignGroup still outperforms all baselines under most evaluation metrics. We owe our superiority to the proposed self-supervised alignment task and our group user jointly optimization strategy. The poor performance of GroupIM and CubeRec verifies the effectiveness of the joint optimization strategy compared with the two-stage optimization strategy.
    \item It is worth mentioning that hypergraph-based methods (HyperGroup and ConsRec) generally perform better than traditional aggregation methods (AGREE), thanks to the excellent representation capabilities of hypergraphs. Also, GroupIM benefits from the self-supervised task to achieve excellent performance. However, our AlignGroup achieves the best performance with excellent motivations that are more in line with the real world.
\end{itemize}

\subsection{Ablation Study (RQ2 \& RQ3 \& RQ4)}
In this section, we conduct exhaustive experiments to examine the effectiveness of various modules in our AlignGroup.

\begin{figure}[h]
    \centering
    \subfigure[HR@5] {
        \label{fig:t}
        \includegraphics[width=0.23\linewidth]{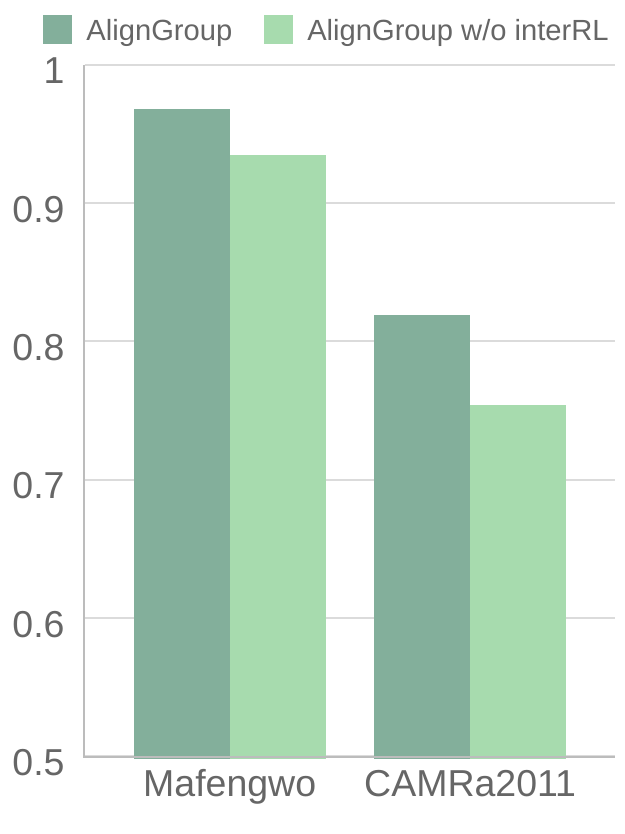}
        }  \hspace{-2.mm}
    \subfigure[HR@10] {
        \label{fig:s}
        \includegraphics[width=0.23\linewidth]{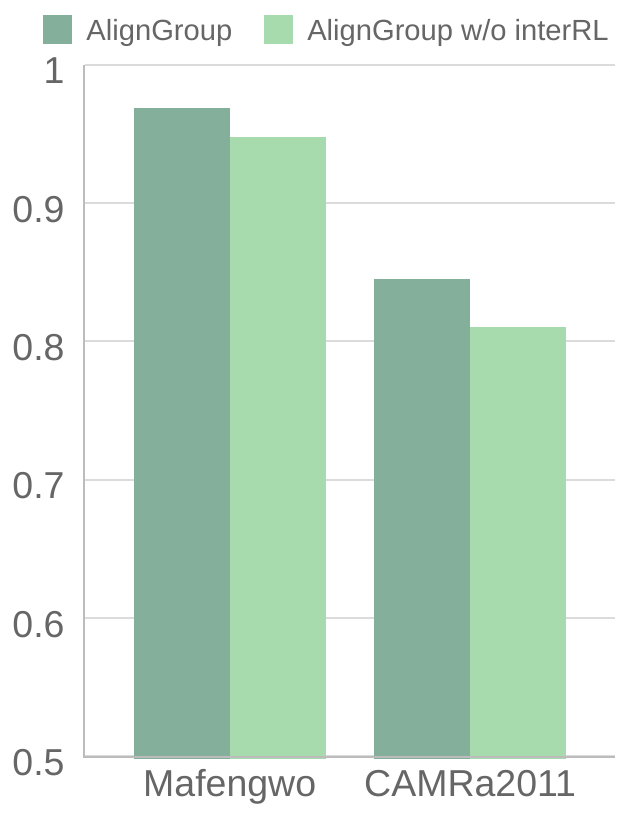}
        }  \hspace{-2.mm}
    \subfigure[NDCG@5] {
        \label{fig:s}
        \includegraphics[width=0.23\linewidth]{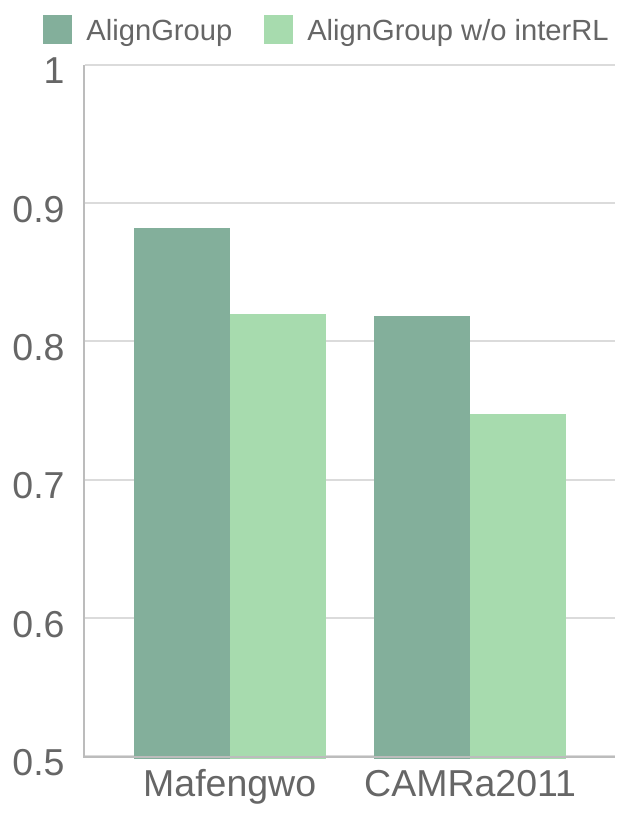}
        }  \hspace{-2.mm}
    \subfigure[NDCG@10] {
        \label{fig:s}
        \includegraphics[width=0.23\linewidth]{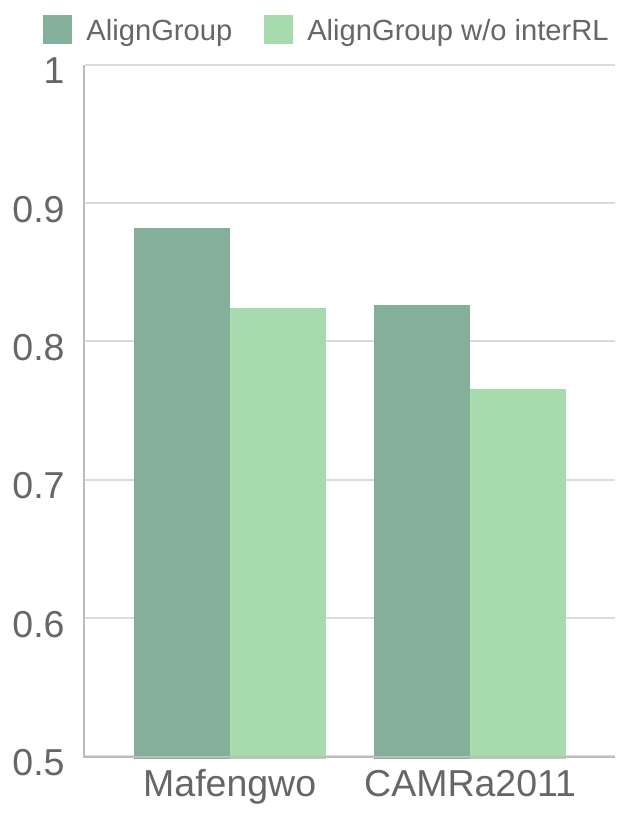}
        }  
        \vskip -0.1in
    \caption{Histograms in dark green show the performance of AlignGroup with both IntraRL and InterRL, and histograms in light green show the performance of AlignGroup w/o InterRL with only IntraRL. The performance metrics are HR@5, HR@10, NDCG@5, and NDCG@10 on both Mafengwo and CAMRa2011 datasets.}   
    \label{fig:RQ2}
     \vskip -0.1in
\end{figure}

\subsubsection{Effectiveness of InterRL (RQ2)}
\label{sec:InterRL}

To verify the effectiveness of the InterRL in our hypergraph neural network, we design the following variant of AlignGroup. 

\textbf{AlignGroup w/o InterRL}: We remove the InterRL part and only retain intra-group relation learning. Formally, we utilize Eq~\ref{eq:5} to compute group representation $\dagger{E}_g$ instead of Eq~\ref{eq:7}.

As shown in Fig~\ref{fig:RQ2}, we can observe that removing InterRL degrades the performance, showing that both intra-group and inter-group relations play a distinct role in group consensus learning.

\subsubsection{Performance of Different Strategies and Scopes of Self-supervised Alignment Task (RQ3)}
\label{sec:ssl}
To analyze the performance of different strategies and scopes of our self-supervised alignment task, we introduce an ablation study. In terms of scope, we refer to considering only the user as "small" and considering both the user and item as "big". Therefore we try to combine two different scopes and two different strategies for computing common preferences:

\textbf{Centroid-small}: We choose \textbf{geometric centroid} as common preferences for each group including \textbf{small} scope members.

\textbf{Centroid-big}: We choose \textbf{geometric centroid} as common preferences for each group including \textbf{big} scope members.

\textbf{Barycenter-small}: We choose \textbf{geometric barycenter} as common preferences for each group including \textbf{small} scope members.

\textbf{Barycenter-big}: We choose \textbf{geometric barycenter} as common preferences for each group including \textbf{big} scope members.

\begin{table}[!ht]
    \centering
\caption{Ablation study on different strategies and scopes of self-supervised alignment task with both group and user recommendation results reported.}
\vskip -0.15in
\label{tab:RQ3}
\resizebox{\linewidth}{!}{
    \begin{tabular}{c|c|c|cccc}
    \hline
         Task&  Dataset&  Variant&  H@5&  H@10&  N@5&  N@10\\
         \hline
         \multirow{8}{*}{Group} & \multirow{4}{*}{Mafengwo} & Centroid-small & \textbf{0.9678} & \textbf{0.9688} & \textbf{0.8817} & \textbf{0.8821} \\
         & & Centroid-big & 0.9075 & 0.9286 & 0.7755 & 0.7825 \\
         & & Barycenter-small & 0.9332 & 0.9401 & 0.8371 & 0.8529 \\
         & & Barycenter-big & 0.8999 & 0.9102 & 0.7493 & 0.7678 \\
         \cline{2-7}
         & \multirow{4}{*}{CAMRa2011} & Centroid-small & \textbf{0.8193} & \textbf{0.8448} & \textbf{0.8180} & \textbf{0.8259} \\
         & & Centroid-big & 0.6572 & 0.8172 & 0.5292 & 0.5812 \\
         & & Barycenter-small & 0.8002 & 0.8285 & 0.7892 & 0.8021 \\
         & & Barycenter-big & 0.6428 & 0.8071 & 0.5246 & 0.5763 \\
          \hline
         \multirow{8}{*}{User} & \multirow{4}{*}{Mafengwo} & Centroid-small & 0.8159 & 0.8643 & 0.7605 & 0.7763 \\
         & & Centroid-big & \textbf{0.8660} & \textbf{0.9037} & \textbf{0.8294} & \textbf{0.8416} \\
         & & Barycenter-small & 0.7889 & 0.8273 & 0.7297 & 0.7382 \\
         & & Barycenter-big & 0.8525 & 0.8891 & 0.8113 & 0.8302 \\
         \cline{2-7}
         & \multirow{4}{*}{CAMRa2011} & Centroid-small & \textbf{0.8116} & \textbf{0.8382} & \textbf{0.8083} & \textbf{0.8167} \\
         & & Centroid-big & 0.6867 & 0.8239 & 0.5962 & 0.6407 \\
         & & Barycenter-small & 0.7884 & 0.8491 & 0.7429 & 0.7601 \\
         & & Barycenter-big & 0.6612 & 0.8091 & 0.5817 & 0.6287 \\
         \hline
    \end{tabular}
    }
     \vskip -0.1in
\end{table}

We report the experimental results in Table~\ref{tab:RQ3}. Selecting the geometric centroid to calculate shared member preferences yields superior results compared to using the geometric barycenter for both group and individual user recommendation tasks across all evaluation metrics. In most cases, the small scope outperforms the big scope. However, for user recommendation tasks, the big scope shows better results on the Mafengwo dataset. We speculate that this is because the Mafengwo dataset has far fewer items than users, and taking items into common preference computation can make the semantics of users and items more similar, which benefits user recommendation tasks. On the other hand, for the CAMRa2011 dataset, where the number of items greatly exceeds that of users, taking items into common preference computation may significantly alter the semantic representation of users, leading to poorer performance in user recommendation tasks.

\subsubsection{Effectiveness of Self-supervised Alignment Task (RQ4)}
We further verify the effectiveness of our self-supervised alignment task through visualization.

\begin{figure}[h]
    \centering
    \vskip -0.1in
    \subfigure[\begin{minipage}{0.35\linewidth}\centering AlignGroup w/o self-supervised alignment task\end{minipage}] {
        \label{fig:t}
        \includegraphics[width=0.5\linewidth]{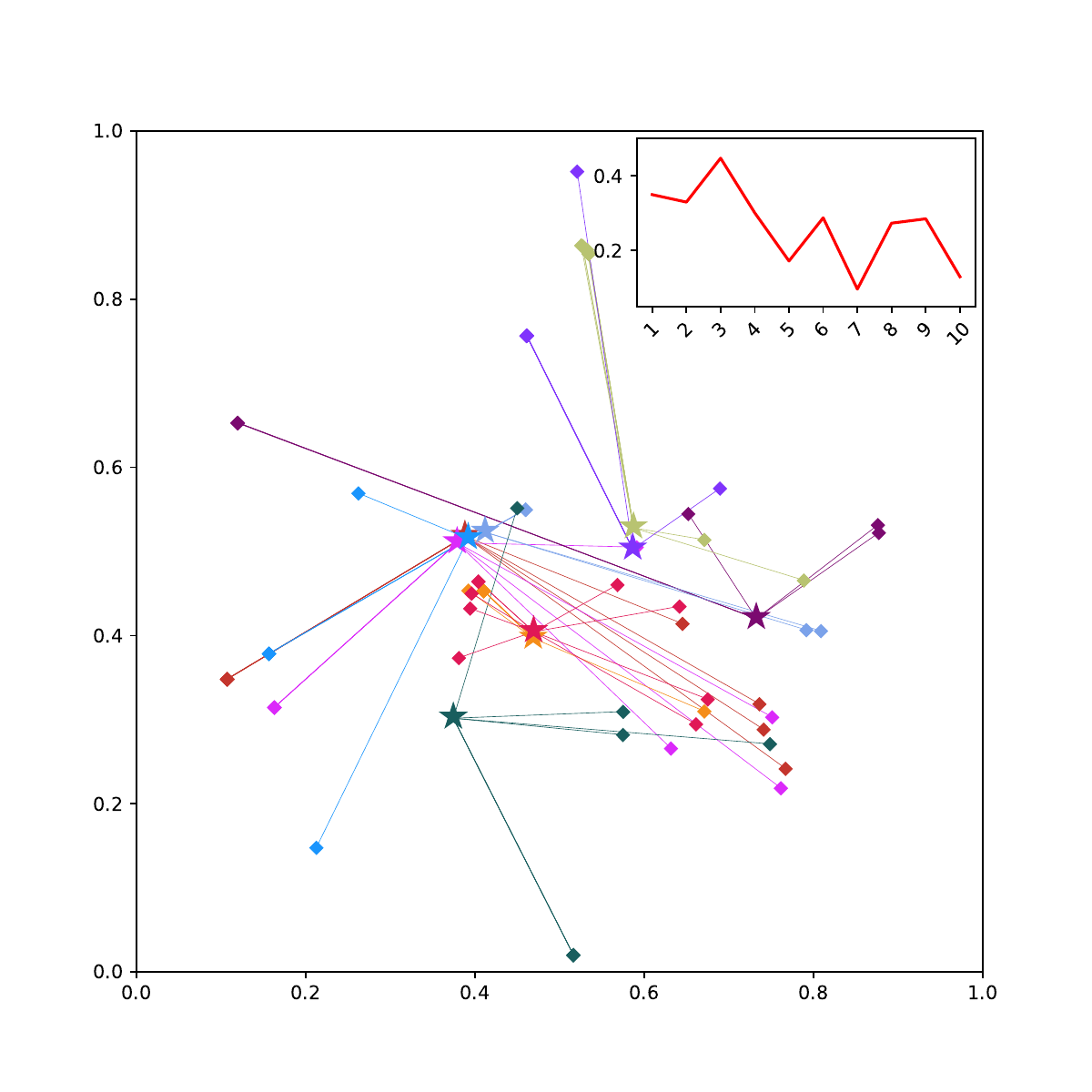}
        } 
        \hspace{-5.mm}
    \subfigure[\begin{minipage}{0.15\linewidth}\centering AlignGroup\end{minipage}] {
        \label{fig:s}
        \includegraphics[width=0.5\linewidth]{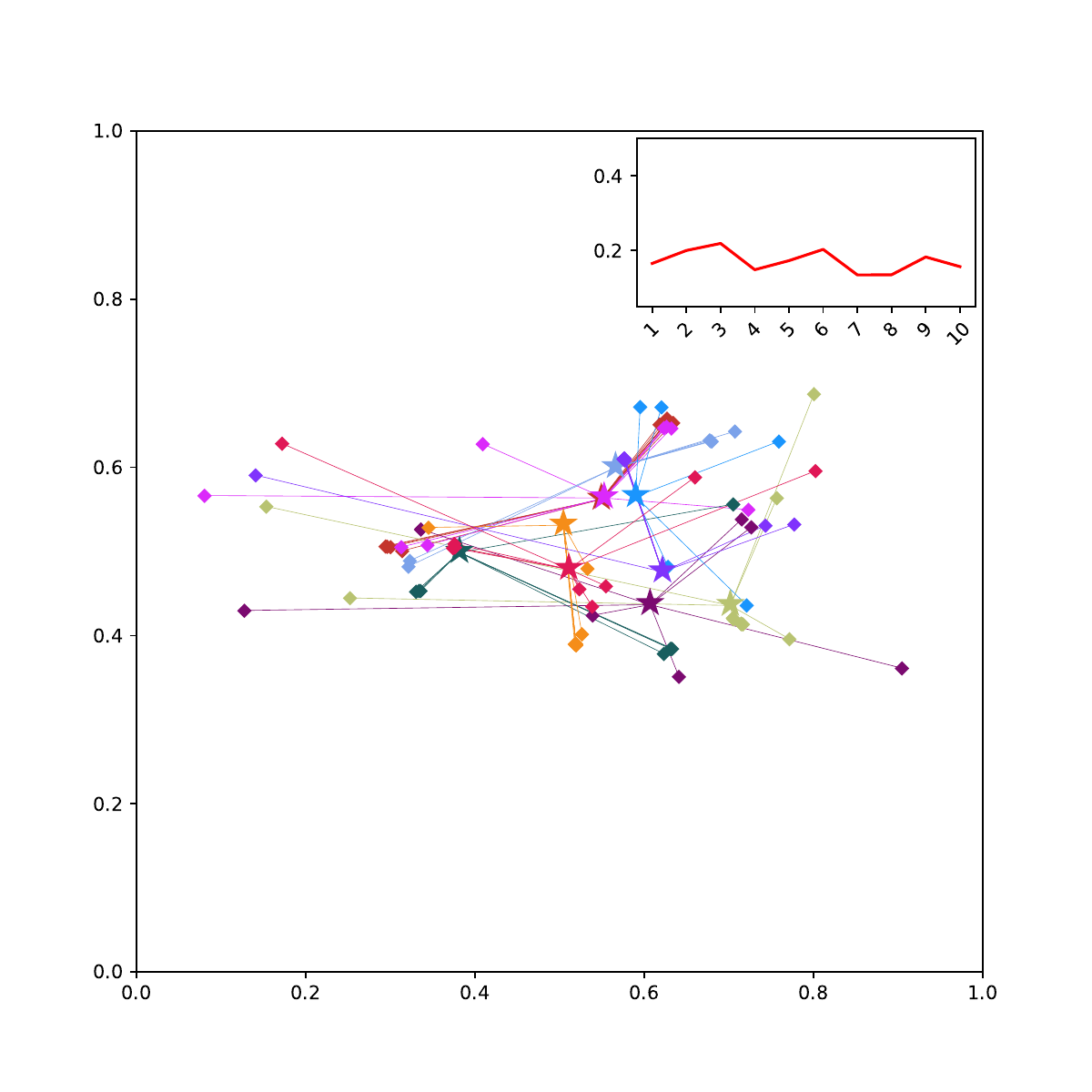} 
        }
    \vskip -0.2in
    \caption{Visualization of (a) AlignGroup w/o self-supervised alignment task and (b) AlignGroup. Pentagram and Prism denote the group consensus and member preferences, respectively. The same color indicates the group consensus and member preferences of the same group. The line graph in the upper right corner represents the average gap between member preferences and group consensus for each of the 10 groups. It shows that AlignGroup in (b) can align member preferences with group consensus much better.} 
    \label{fig:visualization}
    \vskip -0.15in
\end{figure}

Specifically, we randomly sample all users in 10 groups from the Mafengwo dataset. Then, we utilize t-SNE \cite{van2008visualizing} to map their embedding to the 2-dimension space. In Fig~\ref{fig:visualization}, we observe that our self-supervised alignment task effectively reduces the gap between member preferences and group consensus. To show this result more clearly, we average the gap between group consensus and member preferences within the group in the upper right corner of Fig~\ref{fig:visualization}. Our self-supervised alignment task effectively reduces the average gap between group consensus and member preferences.

\subsection{Efficiency Study (RQ5)}
\begin{figure}[h]
    \centering
    \vskip -0.1in
    \subfigure[Mafengwo] {
        \label{fig:m}
        \includegraphics[width=0.5\linewidth]{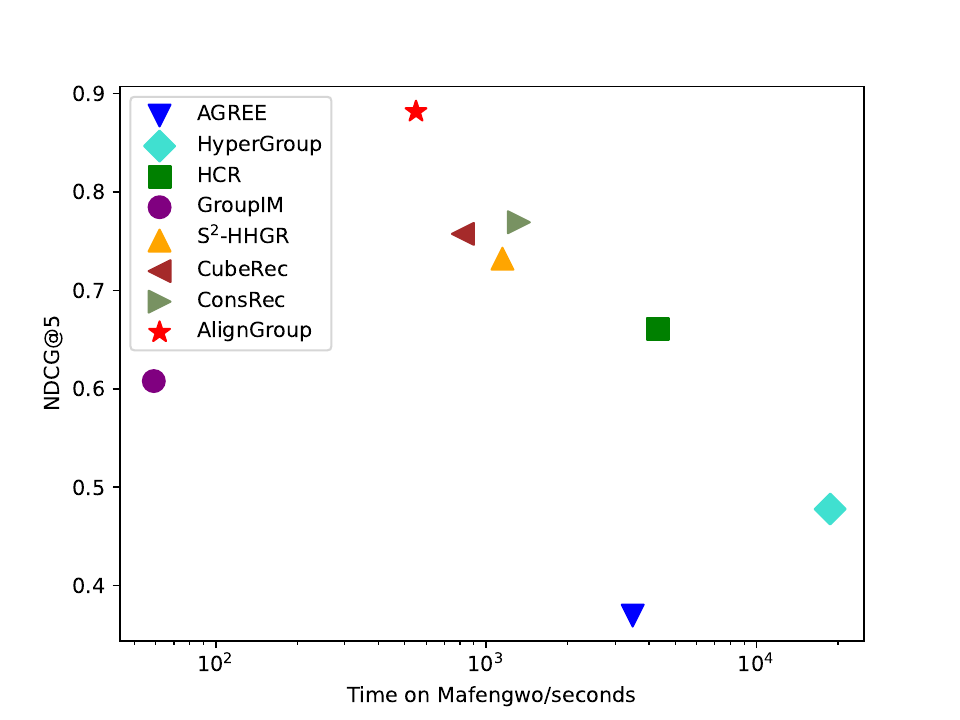}
        }  \hspace{-5.mm}
    \subfigure[CAMRa2011] {
        \label{fig:c}
        \includegraphics[width=0.5\linewidth]{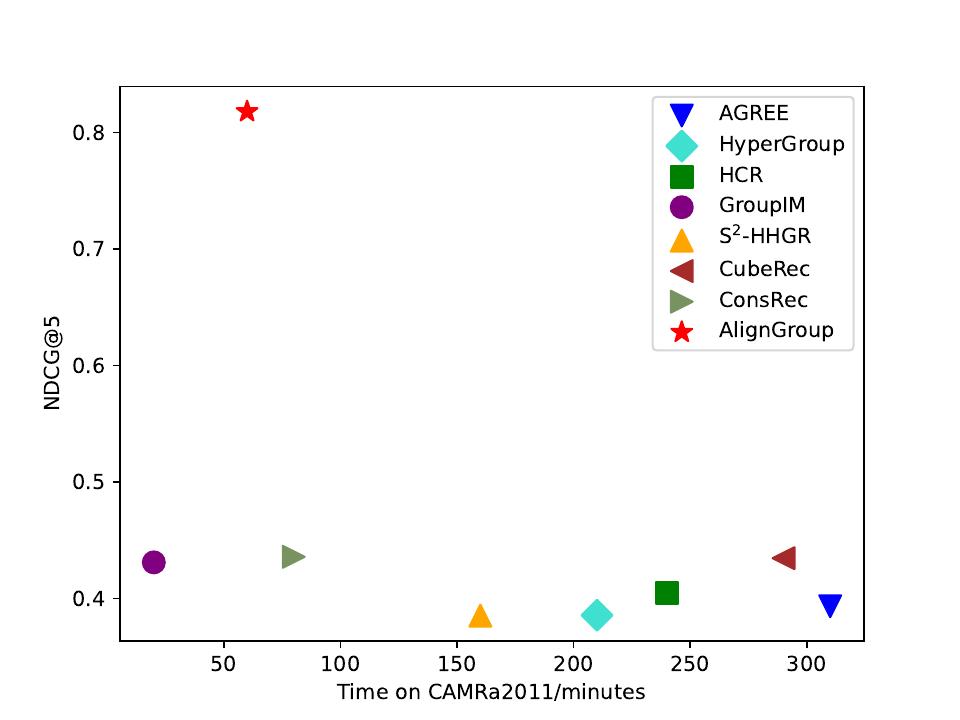}
        }  
    \vskip -0.1in
    \caption{Efficiency Study in terms of NDCG@5 on both Mafengwo and CAMRa2011 datasets.}   
    \label{fig:efficiency}
    \vskip -0.1in
\end{figure}

Following ConsRec \cite{wu2023consrec} settings, we estimate the efficiency of AlignGroup by directly comparing the total running time (including both training and testing) with all baselines. Fig~\ref{fig:efficiency} illustrates the performance (NDCG@5) and runtime (in seconds or minutes) for the group recommendation task on two experimental datasets. Our proposed AlignGroup achieves exceptional efficiency and performance, outperforming all baseline methods in terms of efficiency—with the sole exception of GroupIM—and significantly surpassing all baseline methods in performance.

\subsection{Hyper-parameter Analysis (RQ6)}

\subsubsection{Effect of the Layer Number $L$ of the Hypergraph Neural Network}
\begin{figure}[h]
    \centering
    \vskip -0.1in
    \subfigure[Group recommendation task] {
        \label{fig:s}
        \includegraphics[width=0.48\linewidth]{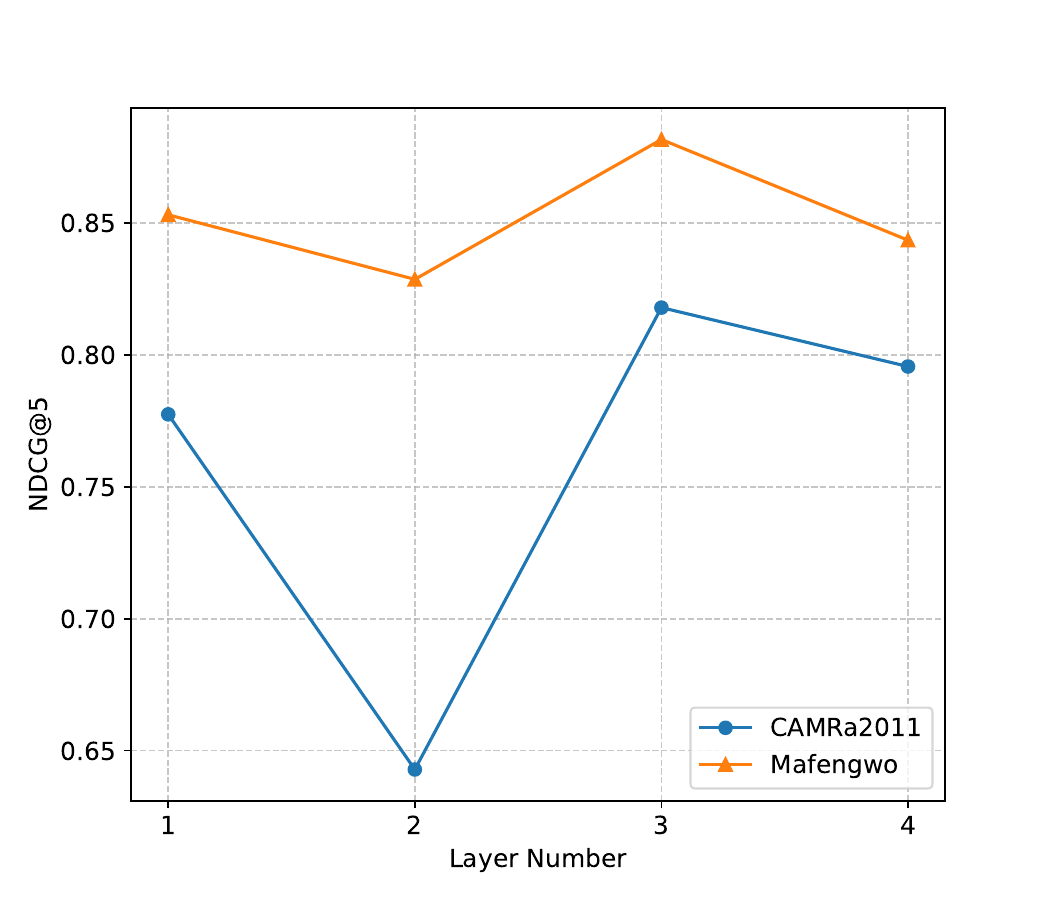}
        }  \hspace{-2.mm}
    \subfigure[User recommendation task] {
        \label{fig:t}
        \includegraphics[width=0.48\linewidth]{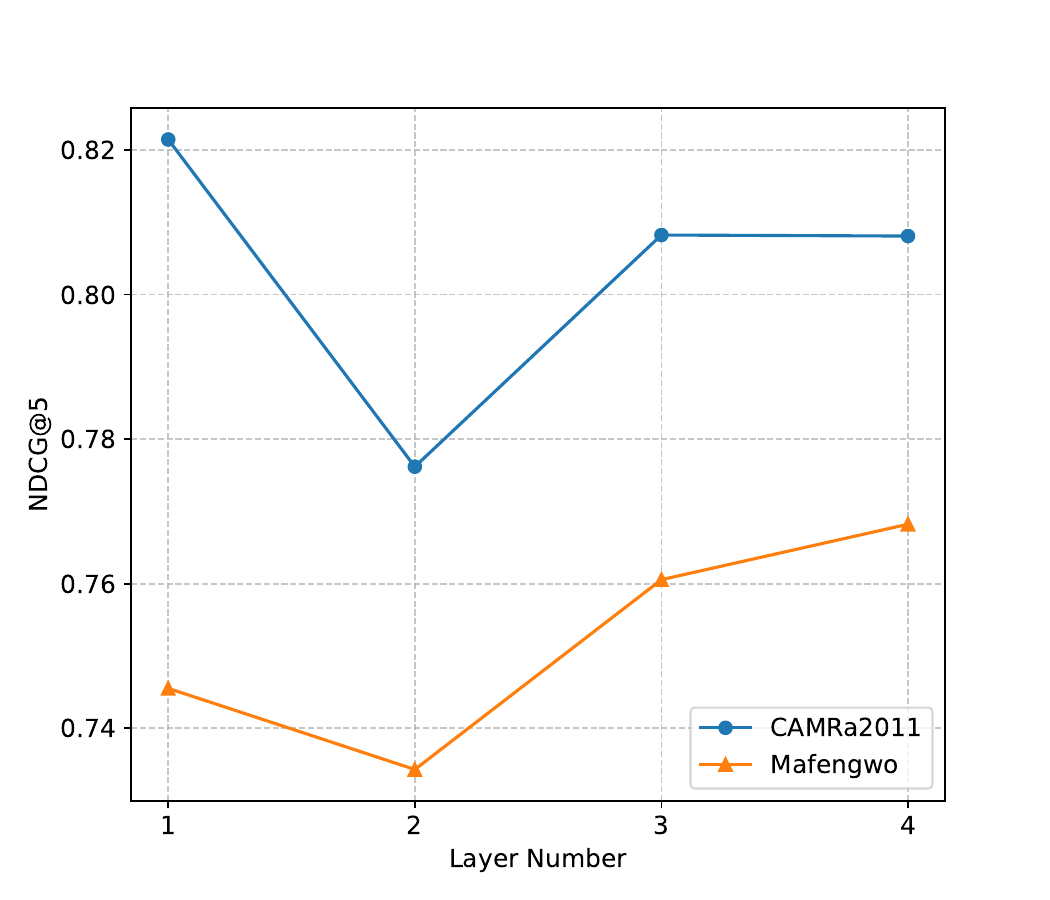}
        }
        \vskip -0.1in
    \caption{Performance of AlignGroup with respect to different layer numbers $L$ of hypergraph neural network.}   \vskip -0.1in
    \label{fig:line}
\end{figure}

Layer number affects the performance of the hypergraph neural network. More layers can aggregate more information from high-order neighbors. However, as the layer number increases, it faces the over-smoothing problem \cite{li2018deeper} where the discriminative of node representations is insufficient. Fig~\ref{fig:line} illustrates the performance trends of AlignGroup with different settings of $L$. Considering both the group recommendation and user recommendation tasks, $L$ = 3 is the optimal setting on both Mafengwo and CAMRa2011 datasets.

\subsubsection{Effect of the Balancing Hyper-parameters $\lambda_{align}$ and $\tau$ of the Self-supervised Alignment Task}
\begin{figure}[h]
    \centering
    \vskip -0.1in
    \subfigure[Group task on Mafengwo] {
        \label{fig:s}
        \includegraphics[width=0.48\linewidth]{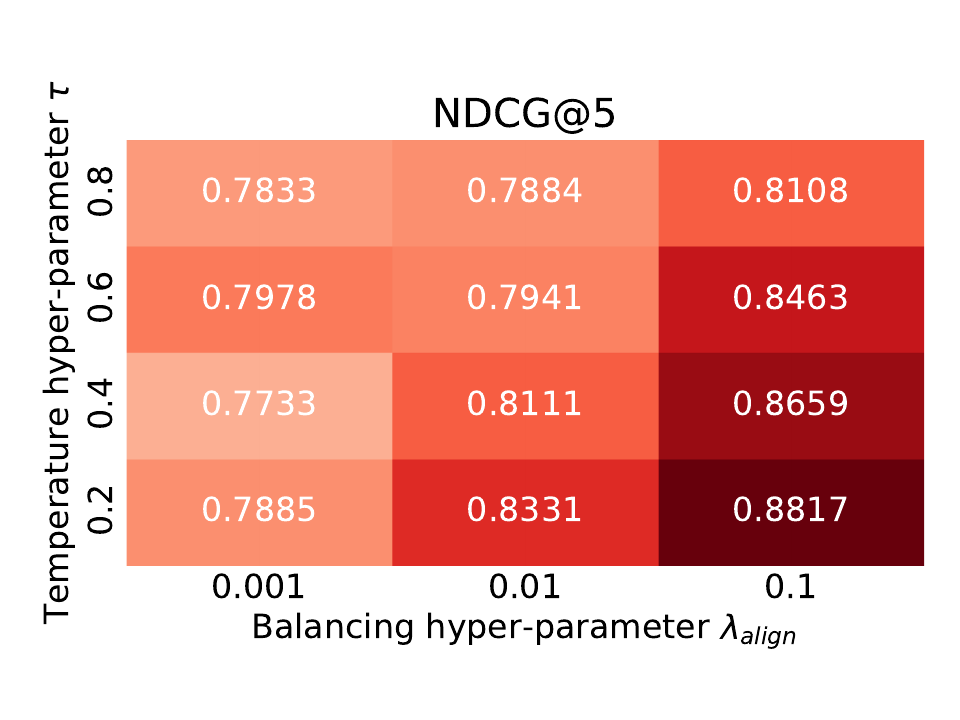}
        }  \hspace{-2.mm}
    \subfigure[User task on Mafengwo] {
        \label{fig:t}
        \includegraphics[width=0.48\linewidth]{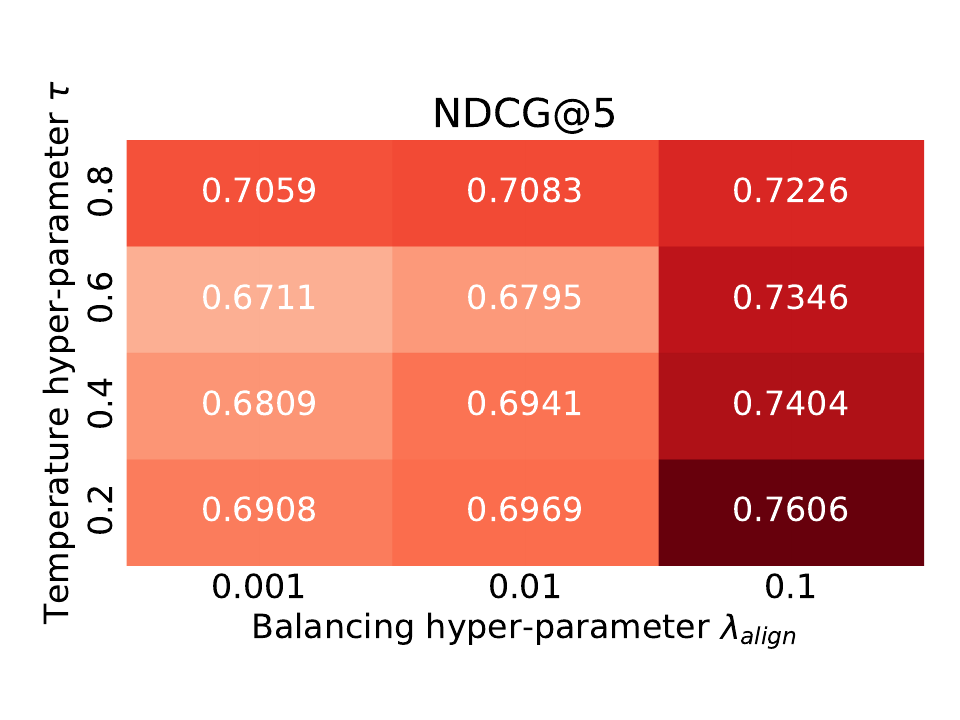}
        }  \hspace{-2.mm}
        \vskip -0.15in
    \subfigure[Group task on CAMRa2011] {
        \label{fig:s}
        \includegraphics[width=0.48\linewidth]{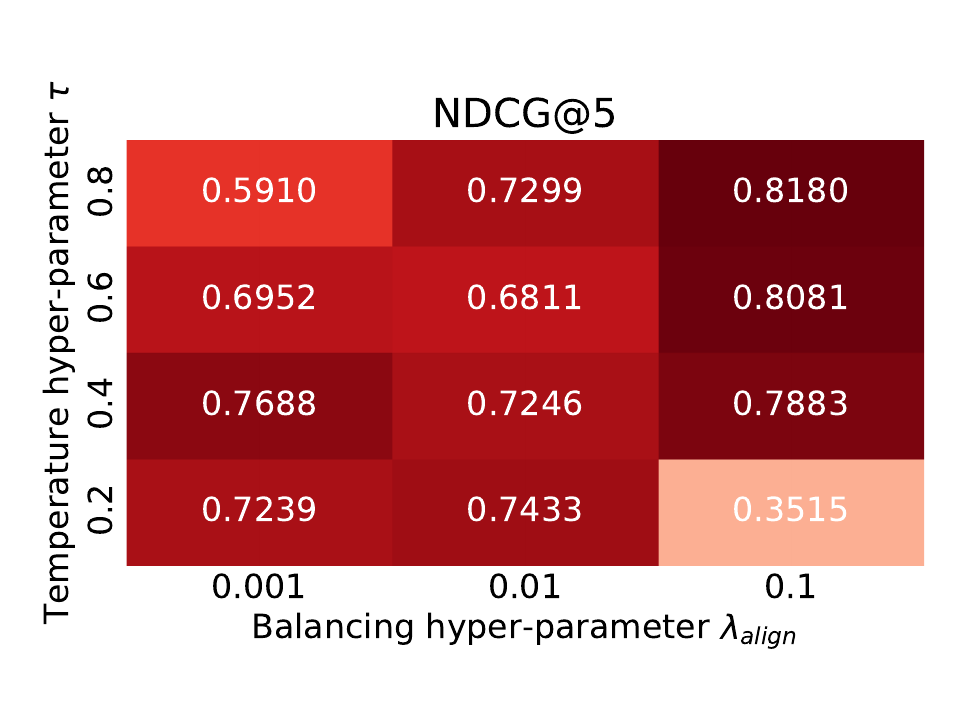}
        }  \hspace{-2.mm}
    \subfigure[User task on CAMRa2011] {
        \label{fig:s}
        \includegraphics[width=0.48\linewidth]{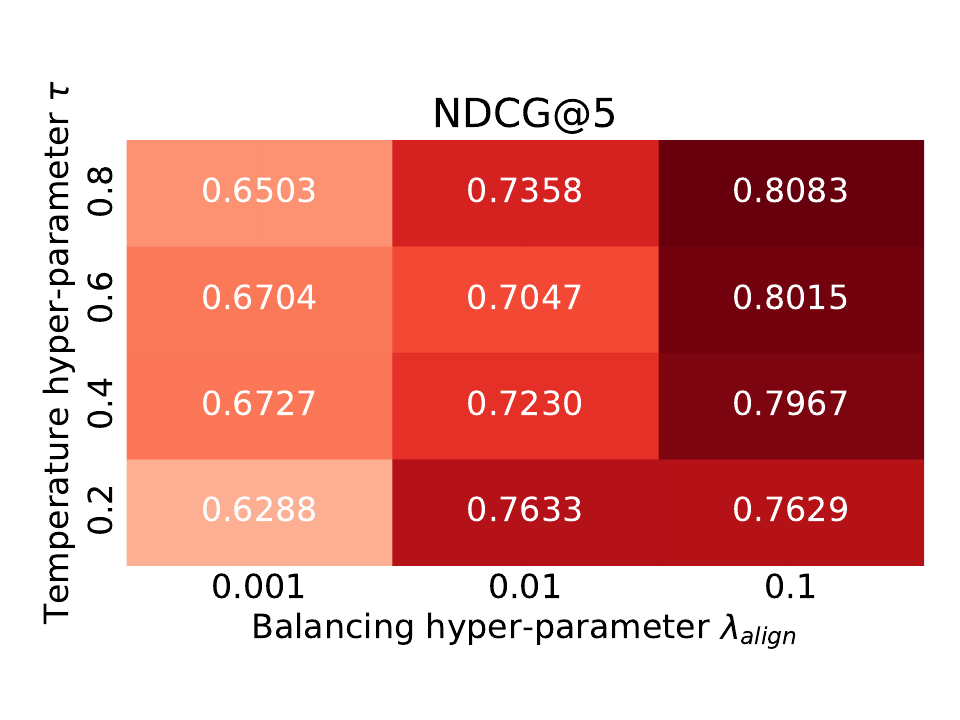}
        }  
        \vskip -0.15in
    \caption{Performance of AlignGroup with respect to different hyper-parameter pairs ($\lambda_{align}$,$\tau$). Darker color denotes better performance of recommendation.} 
    \vskip -0.15in
    \label{fig:heatmap}
\end{figure}

The balancing hyper-parameter $\lambda_{align}$ and the temperature hyper-parameter $\tau$ jointly control the degree of our self-supervised alignment task. We tune the $\lambda_{align}$ from \{1e-1,1e-2,1e-3\}, and $\tau$ from \{0.2, 0.4, 0.6, 0.8\}. Fig.~\ref{fig:heatmap} shows that the best performances for both group recommendation and user recommendation tasks are achieved with $\lambda_{align}$ = 1e-1 and $\tau$ = 0.2 on Mafengwo dataset. On CAMRa2011 dataset, the optimal result is achieved with $\lambda_{align}$ = 1e-1, $\tau$ = 0.8.

\section{Conclusion}
In this paper, we propose a novel group recommendation method, named AlignGroup. It extracts group consensus by learning intra- and inter-group relations and further refining the group decision-making by aligning group consensus and members' common preferences. In particular, AlignGroup effectively captures the group consensus from both intra- and inter-group relations by a well-designed hypergraph neural network. Moreover, AlignGroup proposes a novel and simple self-supervised alignment task to capture fine-grained group decision-making by aligning the group consensus with members' common preferences. Experimental results on two real-world datasets rigorously verify the effectiveness and efficiency of AlignGroup for both group and user recommendation tasks.

\begin{acks}
This work was supported by the Hong Kong UGC General Research Fund no. 17203320 and 17209822, and the project grants from the HKU-SCF FinTech Academy.
\end{acks}


\balance

\end{document}